\documentclass[12pt]{article}
\usepackage[dvips]{epsfig}

\begin{document}

%
\catcode`\@=11
\@addtoreset{equation}{section}
\def\theequation{\thesection.\arabic{equation}}
\catcode`\@=12
\newcommand{\Mcodepath}{{\tt $\sim$uwolff/na/matlab/noncompact }}
\newcommand{\be}{ \begin{equation}}
\newcommand{\ee}{ \end{equation}  }
\newcommand{\bea}{ \begin{eqnarray}}
\newcommand{\eea}{ \end{eqnarray}  }
\newcommand{\benn}{ \begin{equation} \nonumber }
\newcommand{\beann}{ \begin{eqnarray*}}
\newcommand{\eeann}{ \end{eqnarray*}  }
\newcommand{\bi}{\bibitem}
\newcommand{\re}{ \mbox{\rm e} }
\newcommand{\rO}{ \mbox{\rm O} }
\newcommand{\tr}{\mbox{tr}}
\newcommand{\Tr}{\mbox{Tr}}
\newcommand{\D}{\Delta}
\newcommand{\llangle}{\langle\langle}
\newcommand{\rrangle}{\rangle\rangle}
\renewcommand{\floatpagefraction}{0.8}

\evensidemargin0.5cm
\oddsidemargin0.5cm

\newcommand{\del}{\partial}
\newcommand{\sort}{{\rm sort}}
\newcommand{\erfc}{{\rm erfc}}

\renewcommand{\Re}{\mbox{Re}}
\renewcommand{\Im}{\mbox{Im}}
\title{
Dynamical fermions as a global correction
}

\author{
Francesco Knechtli and Ulli Wolff\thanks{
e-mail: knechtli@physik.hu-berlin.de, uwolff@physik.hu-berlin.de} \\
Institut f\"ur Physik, Humboldt Universit\"at\\ 
Newtonstr. 15 \\ 
12489 Berlin, Germany
}
\date{}
\maketitle
\vbox{
\centerline{
\epsfxsize=2.5 true cm
\epsfbox{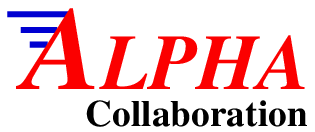}}
}
\vskip 0.5 cm

\vspace*{3cm}

\begin{abstract}
In the simplified setting of the Schwinger model
we present a systematic study
on the simulation of dynamical fermions
by global accept/reject steps that take into account
the fermion determinant.
A family of exact algorithms is developed, which combine
stochastic estimates of the determinant ratio
with the exploitation of
some exact extremal eigenvalues of the generalized problem
defined by the `old' and the `new' Dirac operator.
In this way an acceptable acceptance rate is achieved
with large proposed steps and
over a wide range of couplings and masses.

\end{abstract}
\begin{flushright} HU-EP-03/12\\ SFB/CCP-03-07 \end{flushright}

\thispagestyle{empty}
\newpage


\section{Introduction}

The problem to include fermionic fluctuations in QCD
simulations has been in the focus of interest of the
lattice community for a long time. An overview over
the standard approaches to such simulations can be found
in \cite{Creutz-Book}. About the most obvious idea that
one may have, is the following succession of steps:
First we propose changes of
the gauge field by some efficient algorithm
that fulfills detailed balance with respect to some
suitable pure gauge action. Then we add a
Metropolis accept/reject step. This correction
has to filter the ensemble such that it becomes  governed 
by the desired action including
the fermion determinant, whose change enters the accept decision.
Since the proposal is not guided
by the fermions, one may however fear to get sufficient
acceptance only for tiny rather local changes of the
gauge field and to get an overall very inefficient
algorithm by a succession of such steps.
Alternative approaches, including the presently
most popular hybrid Monte Carlo algorithm
\cite{Duane:1987de,Gottlieb:1987mq} (HMC),
therefore use some sort of stochastic representation
of fermion effects to guide the gauge field evolution at
the expense of introducing additional noise and having to make
many small and expensive steps involving inversions of the
Dirac operator.

While fermionic guidance may prove indispensable for many
lattice simulations, in our opinion
there is also some interest to pursue the direct Metropolis approach.
One reason is the growing interest in Dirac operators, where the
fermions are coupled to smoothed
SU(3)-projected averages of the fundamental gauge fields
\cite{Orginos:1999cr,Hasenfratz:2001hp,Hasenfratz:2000qb,
Bietenholz:2000iy,DeGrand:2000tf}.
Here the dependence of the operators on the fundamental fields to be updated
becomes so complicated, that their change even under infinitesimal moves,
which leads to
the fermionic force, becomes impracticable to evaluate both on paper
and CPU. Metropolis on the other hand requires nothing but a routine
that is able to apply the Dirac operator to given fields. 
A similar situation prevails
with Ginsparg Wilson fermions, for instance in the form of
the overlap formulation \cite{Neuberger:1998fp}.

An additional important motivation for the present investigation 
stems for us from our interest
in simulations of the Schr\"odinger functional with
dynamical fermions \cite{Bode:2001jv,Knechtli:2002vp}. 
Here one is also
interested in larger $\beta$-values implying small physical
volume and approximate validity of perturbation theory.
Then we expect fluctuations of
the determinant to become a small (one-loop) effect.
Our future hope is to develop a Metropolis algorithm for
the determinant whose efficiency in the  large $\beta$ 
limit becomes more similar to pure
gauge simulations based on hybrid overrelaxation. HMC-type
algorithms on the other hand remain rather costly also in this
limit. If such an algorithm can be constructed, it will be interesting
to see, if and where there is a cross-over in efficiency compared to HMC.

Reliable algorithmic optimizations with dynamical fermions
in four dimensional theories are very costly and large scale
projects by themselves.
Therefore --- like other researchers --- we decided to first undertake
a study in two-dimensional QED,
the Schwinger model 
\cite{Lang:1998tm,Gattringer:1997qc,Bardeen:1997bp,Narayanan:1995sv}.
This simplification allows
for clean clinical tests 
using for instance the precise knowledge of the full spectrum of fermionic
operators.  We are of course aware of the risk, that
the smoother gauge fields in this superrenormalizable model may teach us lessons
that do not carry over to QCD. Therefore we plan to soon test
the algorithms derived here in the four dimensional Schr\"odinger
functional.

Other efforts to use the determinant directly in Metroplis steps
have been reported
in \cite{Joo:2001bz} for Wilson fermions and in \cite{Hasenfratz:2002jn}
for staggered fermions with blocked links.
In \cite{Hasenbusch:1998yb} a hierarchical system of acceptance steps has
been tested. Although interesting, we think it is fair to say that
none of these projects has led to a strong competitor for HMC
for standard actions to date.

Our two dimensional study presented here is organized as follows.
In section 2 we set up the notation and our 
lattice formulation of the Schwinger
model. In section 3 we investigate the behavior of the global
Metropolis algorithm with determinants evaluated exactly.
Our main result is contained in section 4 where the stochastic estimation
of determinants is introduced together with a new class of
partially stochastic updates, which is tested in a number of applications
in section 5. After conclusions two appendices follow where we derive
exact formulas for the acceptance rate as a function of the eigenvalues
in an associated generalized eigenvalue problem. The perturbative solution
of this problem is discussed in appendix B.

\section{Model laboratory}
In this section we introduce 
our formulation of
the Schwinger model
discretized as 
two dimensional 
noncompact U(1) gaugefields and Wilson fermions.
Quenched gaugefields are generated by a global heatbath.
We work in lattice units setting the lattice spacing
$a=1$.

\subsection{Formulation of the path integral}
Gauge potentials $A_\mu(x)$ are taken as the primary fields
which are integrated over all real values.
In terms of the field strength
\be
F_{\mu\nu} = \D_\mu A_\nu - \D_\nu A_\mu ,
\label{Fmunu}
\ee
where $\D_\mu$ is the {\em forward} difference, 
the gauge action reads
\be
S_G[A] = \frac14 \sum_{x \mu\nu} F_{\mu\nu}^2 
\label{photonaction}
\ee
with $\mu,\nu=0,1$.
A well-defined path integral on a finite torus of length $L$
yields the pure gauge partition function
\be
Z_G = \int DA \prod_\mu \delta(\sum_x A_\mu)
\prod_x \delta(\D_\mu^* A_\mu) \exp(-S_G[A]),
\label{Aint}
\ee
where $\D_\mu^*$ means the backward difference and $DA=\prod_{x\mu} d A_\mu(x)$.
The $\delta$-functions fix all modes that do not receive damping by $S_G$.
In addition to the gauge degrees of freedom these are two modes
corresponding to constant shifts of $A_\mu$ that we shall come back to.
For later use we abbreviate the normalized full gauge measure as
\be
D\mu(A) = \frac1{Z_G} DA \, \prod_\mu \delta(\sum_x A_\mu)
\prod_x \delta(\D_\mu^* A_\mu) \exp(-S_G[A])
\label{Gaussmeasure}
\ee
and the gauge average as
\be
\langle X(A) \rangle_G = \int D\mu(A) X(A).
\ee

To couple fermions to $A_\mu(x)$ in a gauge invariant fashion
we choose a coupling strength $g$
and form phases
\be
U_\mu(x) = \exp(igA_\mu(x))
\label{UfromA}
\ee
and covariant difference operators
\bea
D_\mu   f(x) &=& U_\mu(x) f(x+\hat{\mu}) - f(x) 
\label{covder}\\
D^*_\mu f(x) &=& f(x) - U^*_\mu(x-\hat{\mu}) f(x-\hat{\mu}).
\label{covderstar}
\eea
Now the Wilson operator reads
\be
D_W = \sum_\mu \frac12 \left\{\gamma_\mu (D_\mu + D^*_\mu) - D^*_\mu D_\mu \right\}
\ee
with some choice of hermitian $\gamma$ matrices.
In terms of 
\be
\gamma_5 = i \gamma_0 \gamma_1
\ee
we have the pseudo-hermiticity property
\be
D_W^\dag = \gamma_5 D_W \gamma_5 \, .
\ee

For our algorithmic study we choose periodic boundary conditions
for the fields that $D_W$ acts on.
Had we allowed  constant components in
$A_\mu$ then we could transform them away by
a  non-periodic gauge transformation.
This would  however modify
the fermion boundary conditions by extra phase factors.
Our constraint may thus be viewed as a definite set of
boundary conditions in imposing a finite volume.

The partition function for $N_f$ flavours of mass $m$ is taken as
\be
Z = \int D\mu(A) \det(D_W+m)^{N_f} .
\label{ZNf2}
\ee
In the following 
we restrict ourselves to the strictly positive case $N_f=2$
analogous to QCD with only light degenerate flavours.

Wilson loops 
constructed from the phases $U_\mu$ decay in the gauge ensemble
with an exact area law.
The string tension
\be
\sigma = \frac{g^2}{2}
\ee
is used to eliminate the dimensionful coupling $g$
in favour of the dimensionless combination
\be
z=\sqrt{\sigma} L \, .
\ee
In the pure gauge theory the limit $L\to\infty$ at fixed $z$
corresponds to a continuum limit at finite physical volume.

\subsection{Generation of gauge fields}

Gauge fields distributed with  $D\mu(A)$ can be generated by a global
heatbath procedure or independent sampling.
A potential $A$ in the Markov chain is followed by $A'$
which, due to
the constraints, can be written as 
\be
A'_\mu = \epsilon_{\mu\nu} \D_\nu^* \phi.
\label{Afromphi}
\ee
The lattice scalar $\phi$ is taken as the Fourier transform
\be
\phi(x) = \frac1L \sum_{p} \tilde{\phi}(p) \exp(i p\cdot x).
\label{phiinx}
\ee
of independent Gaussian random numbers $\tilde{\phi}(p)$
\bea
\tilde{\phi}(0) = 0, \quad \tilde{\phi}^*(p) = \tilde{\phi}(-p) 
\label{phitildenull}\\[0.5ex]
\langle\, \tilde{\phi}^*(p) \, \tilde{\phi}(q) \,\rangle =
\frac{1}{(\hat{p}^2)^2} \delta_{pq} \quad (p\not=0).
\label{phitildecov}
\eea
Momenta are summed over the appropriate Brillouin zone and
$\tilde{\phi}$ depends on them periodically.

It will also be of interest for us to mimic 
smaller update moves $A\to A''$
which are not independent but just fulfill detailed balance with
respect to $S_G$. This can be achieved by taking
\be
A''=cA+sA' \, , \quad 
c=\cos(t\pi/2), \, s=\sin(t\pi/2),
\label{tsteps}
\ee
where the parameter $0<t\le 1$ allows to control the step-size.

In Fig.\ref{spectra} a few complete spectra of the Wilson Dirac operator
are shown for several couplings at $L=8$ in gauge fields 
generated according to (\ref{Gaussmeasure}).
The high degeneracy of the free spectrum is progressively
lifted as $z$ is raised. At the same time the spectrum moves away from the origin.
\begin{figure}[ht]
 \begin{center}
     \includegraphics[width=14cm]{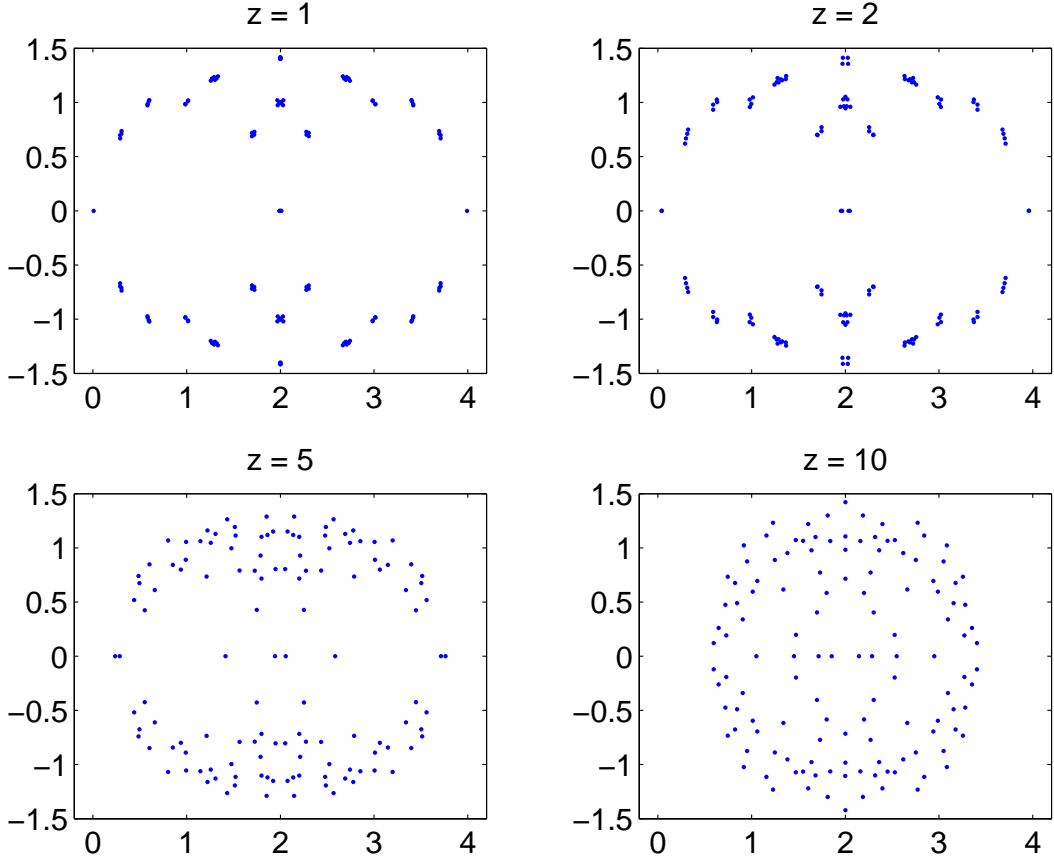}
  \end{center}
 \caption{Typical spectra of $D_W$ at couplings $z=1,2,5,10$}
 \label{spectra}
\end{figure}

For each configuration we define an effective critical mass $m_0$ 
to be the negative spectral gap
\be
m_0(A) = -\min[\Re({\rm spec}(D_W))]
\ee
such that
the smallest real part of all eigenvalues of $D_W+m_0$ just reaches zero.
Its gauge average is taken as a critical value
\be
m_c(z,L) = \langle m_0(A) \rangle_G
\ee
and by $\sigma_c^2(z,L)$ we denote the corresponding variance. 
\begin{table}[htb] 
 \centering
 \begin{tabular}{|c|cc|cc|cc|}
  \hline
       & \multicolumn{2}{c|}{$L=8$} & 
         \multicolumn{2}{c|}{$L=12$} &
         \multicolumn{2}{c|}{$L=16$} \\ \hline
    z  & $m_c/g^2$ & $\sigma_c/g^2$ &  
         $m_c/g^2$ & $\sigma_c/g^2$ &
         $m_c/g^2$ & $\sigma_c/g^2$ \\ \hline\hline
   $1$ & -0.323(3) & 0.088 & -0.382(3) & 0.082 & -0.427(3) & 0.086\\
   $2$ & -0.314(3) & 0.086 & -0.379(3) & 0.084 & -0.420(3) & 0.081\\
   $5$ & -0.289(2) & 0.071 & -0.353(2) & 0.068 & -0.391(2) & 0.069\\
  $10$ & -0.169(1) & 0.020 & -0.262(2) & 0.028 & -0.319(1) & 0.031\\
  \hline 
 \end{tabular} 
\caption{Values for the critical mass and its standard deviation
from 1000 configurations each.}
\label{mctab}
\end{table}
In Table \ref{mctab} some numerical
values are listed. We divide by $g^2$ as suggested by perturbation theory
and obtain numbers that vary only slowly  with $L$ and $z$. For the spectra in
Fig.\ref{spectra}
this implies gaps roughly proportional to $z^2$ in agreement with the four
particular configurations shown.
In our algorithmic study we find it appropriate to use these data
to choose mass values such that our fermions are light and thus
dynamically relevant, but heavy enough to not suffer from
the `exceptional' unphysical modes known to occur with Wilson fermions.

\section{Metropolis with exact determinants}

In this section we use a rather ideal fermion algorithm
that is only available in our two-dimensional model:
global heatbath proposals with respect to the gauge action filtered
through a Metropolis step based on the exact fermionic determinant.
With the cost of computing the determinant by standard linear algebra means 
scaling like $L^{3D}$ this appears prohibitive beyond $D=2$. 
Here on the other hand it will prove to be quite feasible up to medium size lattices
and will provide a rigorous upper bound for the acceptance rates
achievable with stochastic techniques.

\subsection{Exact acceptance rate}

In equilibrium for the ensemble (\ref{ZNf2}) the acceptance
rate $q$ for proposals with the pure gauge global heatbath described
in the previous section
is given by
\be
q = \frac1Z \int D\mu(A) |\det(D_W+m)|^2 \int D\mu(A') \min\left( 1, 
\frac{|\det(D'_W+m)|^2}{|\det(D_W+m)|^2}
\right),
\label{qdet}
\ee
where $A$ and $A'$ enter into $D_W$ and $D'_W$.
In a more symmetric form this reads
\bea
q &=& \frac1Z \int D\mu(A) \int D\mu(A')
\min\left(|\det(D_W+m)|^2 , |\det(D'_W+m)|^2 \right) \nonumber\\[1ex]
&=& 1 - \frac{\int D\mu(A) \int D\mu(A') 
              \left| |\det(D_W+m)|^2 - |\det(D'_W+m)|^2 \right|}
             {\int D\mu(A) \int D\mu(A')
              \left( |\det(D_W+m)|^2 + |\det(D'_W+m)|^2 \right)},
\eea
and $q$ clearly obeys $0\le q \le 1$. Any nontrivial dependence
of the determinant on the gaugefield reduces the acceptance.

To estimate $q$ we generate a large number of $N$ independent gaugefields
with the pure gauge measure
and compute the fermion determinant for each of them.
Let $d_i, i=1, \ldots, N$ be
the resulting successive values of $|\det(D_W+m)|^2)$.
Then $q$ may be estimated by
\be
q \approx \frac{\frac1{N(N-1)}\sum_{i\not=j}\min(d_i,d_j)}
{\frac1{N}\sum_i d_i}.
\label{qest}
\ee
For larger $N$-values the double sum is best evaluated by using
a sorting algorithm
(with cost only $\propto N \ln(N)$, see \cite{Numerical-Recipes}, 
for instance provided in Matlab),
\be
\{ d^<_i \} = \sort(\{ d_i \}),
\ee
where the sequence $\{ d^<_i \}$ consists of the same members as $\{ d_i \}$
but reordered such that $d^<_1 \le d^<_2 \le \ldots \le d^<_N$.
Now the acceptance is written as
\be
q \approx \frac{\sum_i w_i d^<_i}
{\sum_i d^<_i} 
\label{numaccept}
\ee
with the weights
\be
w_i = \frac{2(N-i)}{(N-1)}  \, .
\ee
For the error estimation one of course has to take into account
that the O($N^2$) terms in the numerator of (\ref{qest}) that are effectively
summed by (\ref{numaccept}) are not independent.


It turns out to be very successful to make a Gaussian model
for the distribution of the fermionic action in the gauge ensemble
\be
\nu(E) = < \delta(E - S_F) >_G, \quad S_F = -2\log(|\det(D_W+m)|) ,
\ee
by setting
\be
\nu(E) = \frac1{\sqrt{2\pi}\Sigma} \exp\left( \frac{-(E-E_0)^2}{2\Sigma^2} \right).
\label{Gaussform}
\ee

Once this Ansatz has been made, its free parameters $E_0$ and, more important,
$\Sigma$ can also be estimated numerically from the observed values $d_i$ by
determining mean and variance of $S_F$.
Within the model the acceptance rate then follows,
\be
q = 2 \, \frac{\int_{-\infty}^{+\infty} dE \, \nu(E) \exp(-E)
            \int_{-\infty}^{E} dE' \, \nu(E')}{
            \int_{-\infty}^{+\infty} dE \, \nu(E)\exp(-E)} \, .
\ee
With the expression (\ref{Gaussform}) for $\nu$ one sees that $q$ is independent
of $E_0$ as it should (irrelevance of a constant in $S_F$).
It is convenient to extract $q$ from 
the distribution $\tilde{\nu}(\Delta)$ for the energy difference $\Delta = E'-E$
\be
\tilde{\nu}(\Delta) = \frac1Z \int_{-\infty}^{+\infty} dE \, \nu(E) \exp(-E)
    \int_{\infty}^{+\infty} dE' \, \nu(E') \, \delta(\Delta-E'+E)
\ee
with
\be
Z = \int_{-\infty}^{+\infty} dE \; \nu(E)  \exp(-E) ,
\ee
for which we obtain
\be
\tilde{\nu}(\Delta) = \frac1{2\sqrt{\pi}\Sigma} \exp\left(
-\frac1{4\Sigma^2} (\Delta-\Sigma^2)^2 \right) .
\ee
In terms of $\tilde{\nu}$ we evaluate
\be
q = \int_{-\infty}^{+\infty} d\Delta\; \tilde{\nu}(\Delta) \min(1,\exp(-\Delta)) = 
\frac2{\sqrt{\pi}} \int_{\Sigma/2}^{\infty} du \exp(-u^2) = \erfc(\Sigma/2) .
\label{Gaussaccept}
\ee


We performed a series of (quenched) simulations on $L=8,12,16$ lattices with
$z=1,2,5,10$ (compare Table \ref{mctab}) where we determined and stored the complete
spectra of $D_W$. These data may be used to construct fermion determinants
and acceptance rates for a whole range of masses and in this fashion we
produce Fig.\ref{qvsm}.
\begin{figure}[ht]
  \begin{center}
    \includegraphics[width=12cm]{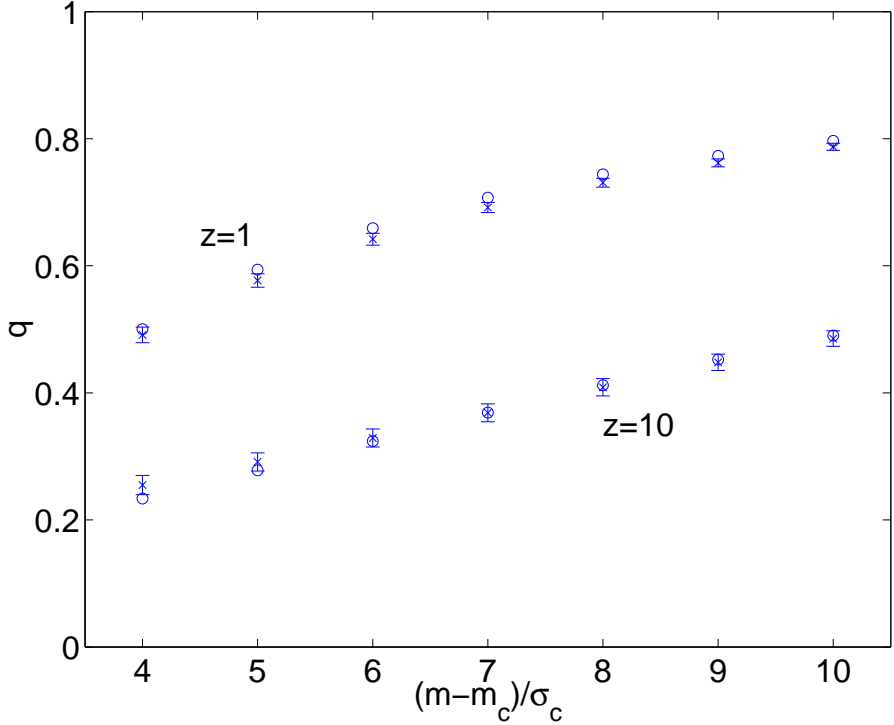}
  \end{center}
\caption{Acceptance rates versus mass for $L=16$. 
Numerical values from (\ref{numaccept})
are shown as crosses and circles correspond to the model 
using (\ref{Gaussaccept}) with the observed $\Sigma$.}
\label{qvsm}
\end{figure}
It demonstrates that in our model, at least with the
exact fermion determinant, a simulation based on global Metropolis
steps is feasible over a wide range of masses. 
Note that this even refers here to maximally large quenched gauge move
proposals. Moreover we verify the validity of the Gaussian model (circles).

\subsection{Enhanced acceptance rate}

So far we have considered the generation of potentials with
the pure gauge action $S_G$ and the incorporation of the
determinant in an accept/reject step. The same ensemble may also be 
produced by a different split of the total action and it is
conceivable that this leads to an enhanced acceptance rate.
It corresponds to a simple version of ultraviolet filtering
\cite{Hasenbusch:1998yb,deForcrand:1998sv,Duncan:1999xh}
by modifying the fermion action just by the plaquette term.
Apart from the gauge action contained in the measure $D\mu(A)$
we include another component
\be
Z = \int D\mu(A) \exp[(1-\alpha^2)S_G] |\det(D_W+m)|^2
\ee
which in simulations is combined with the determinant in the Metropolis test.
By rescaling $A_\mu$ one easily sees that this ensemble
is equivalent to the standard
form of the gauge action with an
effective  coupling of strength $g/\alpha$
in (\ref{UfromA}).
Whenever
the highest acceptance, for fixed $g/\alpha$,
is reached at  $\alpha\not=1$ the extra term
has enhanced the acceptance and is hence advantageous.

In simulations we perform the expensive evaluation of the
determinant
with one fixed value of $g$ in (\ref{UfromA})
and then compute the acceptance for
many values $\alpha$ in the additional term.
In this way we construct lines
in a graph of $q$ versus $z$, where $z$ refers to the effective coupling.
With a number of such lines, we shall see which one gives the highest
acceptance for a $z$ at which we wish to simulate.

The Gaussian model generalizes by
assuming a joint Gaussian distribution for $S_G,S_F$ with
a $2\times 2$ covariance matrix given by connected correlations 
\bea
< S_G S_G >_G^c &=& v_{GG} = \frac12 (L^2-1)\\
< S_F S_F >_G^c &=& v_{FF} \\
< S_G S_F >_G^c &=& v_{GF} = v_{FG} \, ,
\eea
of which the first one is trivial due to the Gaussian gauge action.

Instead of $S_F$ alone, acceptance is now controlled by the combination
$E=S_F+(\alpha^2-1)S_G$ whose variance $\Sigma^2$, given by
\be
\Sigma^2 = (\alpha^2-1)^2 v_{GG} + v_{FF} +2(\alpha^2-1) v_{GF}
\label{SigmaGauss}
\ee
may now be used in (\ref{Gaussaccept}) to estimate $q$.
Within the model, we can easily vary $\alpha$ continuously.

In Fig.\ref{iarates} we see a number of approximate acceptance trajectories
constructed in this way for $L=16,m=-0.1$.
For each of them 1000 gauge configurations were used
and exponentiated in the determinants with coupling $\bar{z}$ corresponding 
to the crosses. The parameter $\alpha$ within the Gaussian model was then varied
in the interval $[0.8,1.2]$ producing the acceptance trajectories as functions
of $z=\bar{z}/\alpha$.
We  clearly see that ultraviolet filtering pays off. At weak coupling
the acceptance is high without it, and correspondingly less can be gained.
\begin{figure}[ht]
  \begin{center}
    \includegraphics[width=12cm]{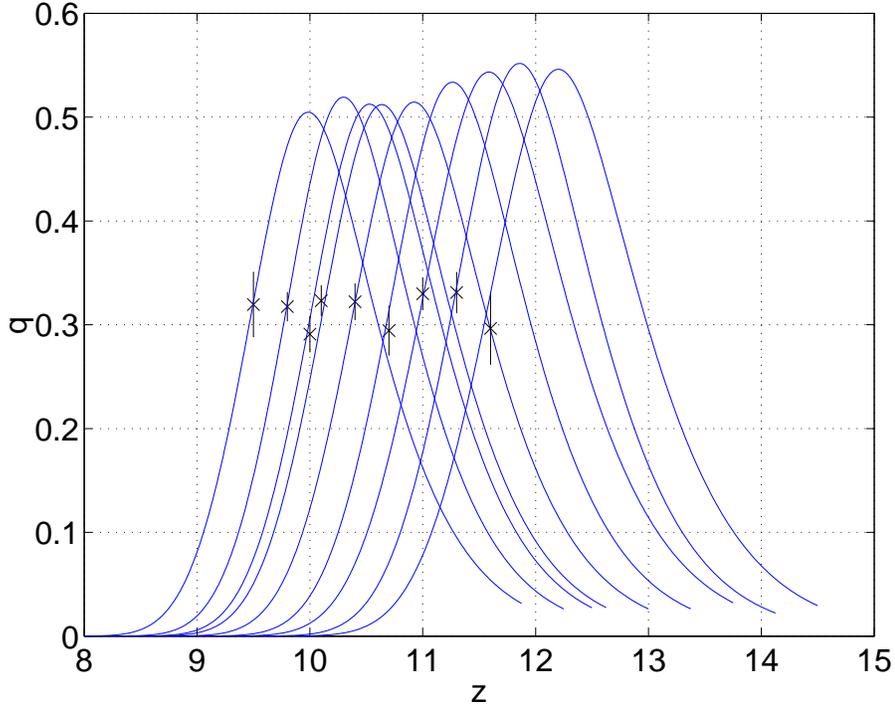}
  \end{center}
\caption{Enhanced acceptance rate in the Gaussian model for $L=16,m=-0.1$.
The crosses correspond to $\alpha=1$.
Each of them extends to a curve
according to (\ref{Gaussaccept}), (\ref{SigmaGauss}), 
as $\alpha$ is varied in the
interval $[0.8,1.2]$.}
\label{iarates}
\end{figure}
For the case with the cross at $\bar{z}=10$ we now confront the Gaussian predictions
for the $\alpha$-dependence of $q$ with numerical values as in (\ref{qest})
where now $d_i=\exp[-S_F-(\alpha^2-1)S_G]$ enters.
This is shown in Fig.\ref{iarateexact} for an ensemble of 1000 gauge potentials
and confirms the model also in this more general setting.
\begin{figure}[ht]
  \begin{center}
    \includegraphics[width=12cm]{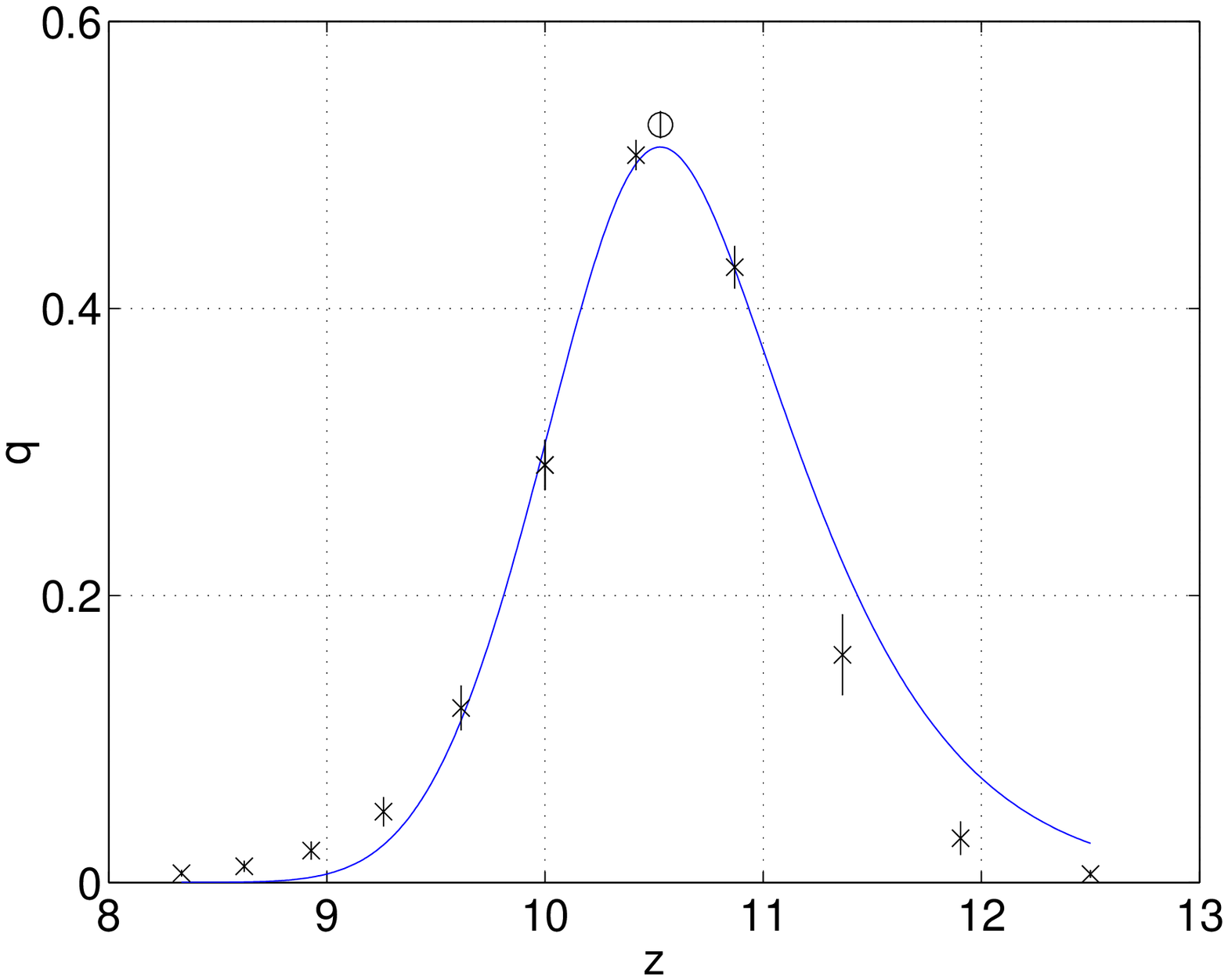}
  \end{center}
\caption{Acceptance rates for $L=16,m=-0.1$.
The curve corresponds to the Gaussian model based on the simulation $\bar{z}=10$,
other data with errors
are direct Monte Carlo checks. The circle is at $\alpha=0.950$
that minimizes (\ref{SigmaGauss}).}
\label{iarateexact}
\end{figure}

\section{Partially stochastic global acceptance steps}

Metropolis steps based on the availability of
exact fermion determinant ratios
are of conceptual interest but hardly lead to efficient
algorithms on large lattices and in four dimensions.
An approximate stochastic estimation that nevertheless
maintains exact detailed
balance can be defined and at first seems more promising 
\cite{Hasenfratz:2002jn,Hasenfratz:2002pt,Hasenfratz:2002vv}.
We found however that this can easily
lead to negligibly small acceptance at smaller masses.
Therefore we developed a more general 
Metropolis correction step
based on an only \underline{p}artially \underline{s}tochastic
estimation of \underline{d}eterminant changes (PSD)
which is complemented by a few exact eigenvalues
[see \cite{Frezzotti:1998yp} for related ideas in connection with
reweighting corrections to the polynomial hybrid Monte
Carlo algorithm].
In this section we shall thus derive a family
of (exact) algorithms which contains those based on exact and on fully stochastic
determinants as extremal cases between which we interpolate.

\subsection{Stochastic estimation of fermion determinant ratios}

The stochastic estimation of the determinant reduces the problem
to solving linear equations with the fermion matrix at the expense of
statistical errors on top of the main Monte Carlo process.
The fundamental formula is
\be
\int D[\eta] \rho(M\eta) = |\det(M)|^{-2}.
\label{stochdet}
\ee
Here we integrate over a complex valued spinor field with the measure
\be
D[\eta] = \prod_{x\alpha} \frac{d\Re(\eta) d\Im(\eta)}{\pi}
\ee
for each site $x$ and spinor component $\alpha$.
For the normalized probability distribution $\rho$,
\be
\int D[\eta] \rho(\eta) = 1
\ee
we often take a Gaussian\footnote{The sum over $x,\alpha$
is included in the scalar product $\eta^\dag \eta$
here and in the following.}
\be
\rho(\eta) = \exp\left[-\eta^\dag \eta \right] \, ,
\ee
but keep the formulas more general where we can.
The determinant  appears as a Jacobian
in (\ref{stochdet}) for arbitrary $\rho$.
We may also write an unbiased estimate
of the determinant as
\be
|\det(M)|^{-2} = \left< \frac{\rho(M\eta)}{\rho(\eta)} \right>_{\eta}
\ee
with an average $< . >_{\eta} $ over the random field $\eta$ only.

For an
acceptance step from a given field $A$ to a proposed new $A'$ with associated
operators $D_W, D_W'$ we now form the ``ratio operator''
\be
M = (D_W'+m)^{-1}(D_W+m),
\label{Maccrej}
\ee
and in terms of this matrix we could stochastically accept with the probability
\be
w_0(A,A') = \min[1,\rho(M\eta)/\rho(\eta)] \, ,
\label{w0stocc}
\ee
where the dependence on $\eta$ and the choice of $\rho$ is left implicit.
For the reverse transition,
$A\leftrightarrow A'$, we find $M \leftrightarrow M^{-1}$.
Therefore
\be
\frac{\left< w_0(A,A') \right>_{\eta}
}{
      \left< w_0(A',A) \right>_{\eta}} = 
\frac{\int D[\eta] \min[\rho(\eta),\rho(M     \eta)]
}{
      \int D[\eta] \min[\rho(\eta),\rho(M^{-1}\eta)]} = 
|\det(M)|^{-2}
\label{DBstocc}
\ee
shows detailed balance. The last equality follows
by changing variables $\eta\to M\eta$ in the integral in the numerator.
These steps constitute the fully stochastic algorithm 
that we are going to
generalize to PSD\footnote{
It would most probably be advantageous to use a preconditioned operator
here. In this study of principles we however avoid this complication.
}.

The acceptance rate is always smaller than (\ref{qdet}) due to the inequality
\bea
\int D[\eta] \min[\rho(\eta),\rho(M\eta)] &\le&
\min\left[
\int D[\eta] \rho(\eta)  ,  \int D[\eta] \rho(M\eta)
\right] \nonumber\\
&=&\min(1,|\det(M)|^{-2})
\eea
Hence, the exact acceptance rate in this context is something like the ideal
``Carnot'' efficiency which we cannot reach but which we may 
also not want to miss by too much.
The question may arise, if the acceptance rate may be increased by averaging over
several random $\eta$ fields\footnote{
This would be possible, if the determinant were incorporated by a
reweighting instead of an acceptance step \cite{Frezzotti:1998eu}.
}.
If we average {\em under} the $\min$ function,
$N_\eta=1$ seems to be the only finite value
for which detailed balance can be shown. Averaging outside of $\min$ seems correct but would not
raise the average acceptance.

An expression for the stochastic acceptance rate 
with distribution $\rho$
is given in analogy
to (\ref{qdet}) by the integral
\bea
q_0 &=& \frac1Z \int D\mu(A) |\det(D_W+m)|^2 \int D\mu(A') \;
\langle w_0(A,A')\rangle_\eta \label{qstoch} \\
&=& \frac1Z \int D\mu(A) |\det(D_W+m)|^2 \int D\mu(A') 
\int D[\eta] \min[\rho(\eta),\rho(M     \eta)]
 \nonumber\\
&=& \frac1Z \int D\mu(A) \int D\mu(A')
\int D[\eta] \min[\rho((D_W+m)^{-1}\eta),\rho((D'_W+m)^{-1}\eta)]
\nonumber
\eea
A naive Monte Carlo  estimation of the last expression for
$q_0$ with $\rho=\exp(-\eta^\dag\eta)$
is not practical
due to the very strong fluctuations of the integrand.
We shall however be able to perform the $\eta$-integrations exactly
in this case, which may be viewed as the construction of an improved
estimator (same mean value, smaller variance)
for $q_0$ in terms of generalized eigenvalues.

We work out the dependence
of 
\be
\langle w_0 \rangle_\eta = \int D[\eta] \min[\rho(\eta),\rho(M     \eta)]
\ee
on the spectrum $\{\lambda_i\}$ of $M^\dag M$
with $i=1,\ldots,n=2L^2$ for the choice $\rho=\exp(-\eta^\dag\eta)$.
Performing the above integration in the basis
of  orthonormal eigenvectors of  $M^\dag M$ with components $z_i$ we find
\be
\langle w_0 \rangle_\eta = 
\prod_i \left(\int \frac{d\Re z_i d\Im z_i}{\pi} \right)
\min\left[\exp(-\sum_i |z_i|^2)\, ,\, \exp(-\sum_i \lambda_i |z_i|^2)\right].
\ee
Changing to polar variables in all the complex planes we get
\be
\langle w_0 \rangle_\eta = \prod_i \left(\int_0^\infty d u_i \right)
\min\left[\exp(-\sum_i u_i)\, ,\, \exp(-\sum_i \lambda_i u_i)\right].
\label{w0polar}
\ee
In appendix \ref{Qformula} this integral is evaluated exactly yielding
\be
\langle w_0 \rangle_\eta = \sum_{i} \; \min(1,1/\lambda_i)
\prod_{j\not=i} \; \frac{\lambda_i-1}{\lambda_i-\lambda_j} \, ,
\label{Qrhoinlambda}
\ee
where the special case of $S$ being the empty set
suffices here (compare (\ref{Qrhouint})).

It is clear from (\ref{w0polar}) that eigenvalues $\lambda_i=1$ are
irrelevant for the acceptance. Approaching this limit for one of them
in (\ref{Qrhoinlambda}) one indeed finds it to drop out and one is left
with the formula for $n-1$ eigenvalues. If we now consider 
as an example the case
of $\lambda_2 \ldots \lambda_{n-1}$ differing from one only negligibly 
and only one remaining pair with $\lambda_n \gg 1 > \lambda_1$
then we find a small acceptance
\be
\langle w_0 \rangle_\eta \approx \frac{2-\lambda_1}{\lambda_n} \, .
\ee
It remains small even for $\lambda_1 \lambda_n =1= \det(M^\dag M)$,
when we have  100\% non-stochastic acceptance. 
This simple example demonstrates how the stochastic acceptance rate degrades
if a determinant ratio of order unity arises from compensations
between the eigenvalues of the squared ratio operator $M^\dag M$.
We are hence motivated to have a closer look at such spectra in our model.

\subsection{Spectrum of random quotients of Dirac operators}

In practice we find the spectrum $\{\lambda_i\}$ of $M^{\dagger}M$ by solving
the generalized eigenvalue problem
\be
 (D_W+m)(D_W+m)^{\dagger}\chi = 
 \lambda (D_W^{\prime}+m)(D_W^{\prime}+m)^{\dagger}\chi \,.
\label{genEV}
\ee
A general observation about the spectrum of $M^{\dagger}M$,
at least for not too strong coupling,  is that there
are two low and two high almost degenerate eigenvalues
separated from the remaining ones.
In the bulk of the spectrum the
eigenvalues are close to one which is
easy to understand, since all eigenvalues would be exactly one if the two gaugefields
entering into $M$ would be equal.

A few experiments show that
the separation of the extremal eigenvalues 
becomes more pronounced as the mass is lowered.
In the same limit and at small coupling they are
of the order $g^2$ and $g^{-2}$ respectively.
\begin{figure}[ht]
 \begin{center}
    \includegraphics[width=12cm]{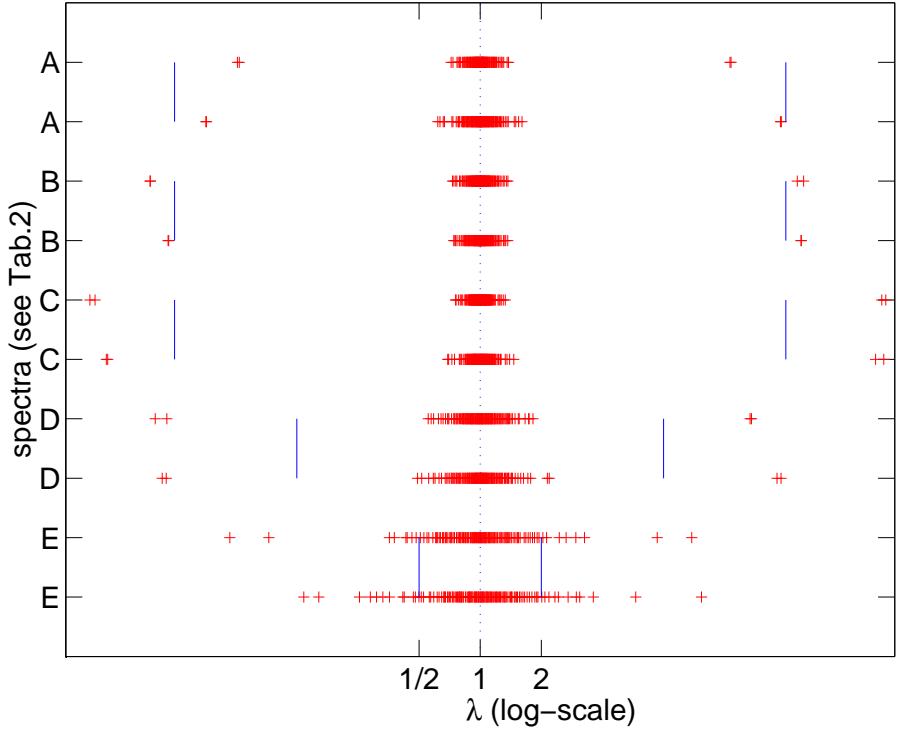}
 \end{center}
 \caption{Typical complete generalized spectra $\{ \lambda_i \}$
 for values $m,z$ found in Tab.\ref{qStab}.
 Vertical bars show the scales $g^2$ and $g^{-2}$.
 \label{specs}}
\end{figure}
This is illustrated in Fig.\ref{specs}
by the example of the quenched
spectra on $L=8$ for several combinations $m,z$.
They are labelled by A--E with the corresponding values
contained in Tab.\ref{qStab}. in Sect.~\ref{qforPSD}.
The plot shows the complete generalized spectra
on a logarithmic scale. For the spectra labelled by A--D
one clearly sees that the bulk is
between 1/2 and 2 with two eigenvalues close to $g^2,g^{-2}$ on either
side, while for the more strongly coupled case E this segregation is lost.

To understand how such a spectrum arises, we consider two hermitian operators
$H$ and $H^{\prime}$ and the generalized Ritz functional (or Rayleigh quotient)
\be
 \mu(\psi) = \frac{\psi^{\dagger}H\psi}{\psi^{\dagger}H^{\prime}\psi} \,.
 \label{ritz}
\ee
The vectors $\psi_{\star}$ for which $\mu(\psi)$ is stationary satisfy the
generalized eigenvalue equation
$H\psi_\star=\lambda H^{\prime}\psi_\star$ with 
$\lambda=\mu(\psi_{\star})$. In particular
the smallest generalized eigenvalue $\lambda_{\rm min}$ has the property
\be
 \mu(\psi) \ge \lambda_{\rm min} \quad \mbox{for all $\psi$} \,.
 \label{minritz}
\ee
In our case $H=(D_W+m)(D_W+m)^{\dagger}$ and
$H^{\prime}=(D_W^{\prime}+m)(D_W^{\prime}+m)^{\dagger}$. The typical
situation is encountered
if we set $m=0$ and assume that $H$ and $H^{\prime}$
are positive for nonvanishing coupling (note that $m_c < 0$).

Instead of considering directly the generalized eigenvalue problem 
(\ref{genEV}) we use the relation $H=D_WD_W^{\dagger}=(D_W\gamma_5)^2$
and first study the
spectrum 
\be
D_W\gamma_5 \, \psi = \kappa \psi
\ee
in perturbation theory.
By expanding
$U_\mu=1+igA_\mu-g^2A_\mu^2/2+\mathcal{O}(g^3)$
we obtain an expansion of the Wilson Dirac operator
\be
 D_W =  (D_W^{(0)} + gD_W^{(1)} + g^2D_W^{(2)} + \rO (g^3)) \,.
\ee
We are interested in the perturbation of the eigenvalue $\kappa^{(0)}=0$
of the free operator $D_W^{(0)}\gamma_5$.
It is doubly degenerate with spatially constant free eigenfunctions
$\psi_\pm^{(0)}$.
We set up expansions
\begin{eqnarray}
 \kappa_\alpha &=& g\kappa^{(1)}_\alpha+g^2\kappa^{(2)}_\alpha+
 \rO(g^3) \, , \\
 \psi_\alpha &=& \psi^{(0)}_\alpha+g\psi^{(1)}_\alpha+
 g^2\psi^{(2)}_\alpha+\mathcal{O}(g^3) \,, \quad \alpha=1,2 \,.
\end{eqnarray}
The perturbation $D_W^{(1)}$ is proportional $A_\mu(x)$
which has no zero momentum component. Hence matrix elements
of this operator vanish between states of equal momentum
and in particular between $\psi_\pm^{(0)}$.
This implies the vanishing of $\kappa^{(1)}_\alpha$.
Suitable zeroth order eigenfunctions $\psi^{(0)}_\alpha$ 
are linear combinations of $\psi_\pm^{(0)}$ which we determine later.
The first order equation for the eigenvectors is
\be
 D_W^{(0)}\gamma_5\psi_\alpha^{(1)} = -D_W^{(1)}\gamma_5\psi_\alpha^{(0)} \,,
 \quad \alpha=1,2\,. \label{firstorder}
\ee
In second order the $2\times2$ matrix
\be
 \psi^{(0)}_k \left[
 D_W^{(2)}-D_W^{(1)}D_W^{(0)-1}D_W^{(1)}\right]\gamma_5 \psi^{(0)}_l\,,
 \quad k,l=\pm \,
\ee
has to be diagonalized. Here the eigenvectors $\psi^{(0)}_\alpha$
get determined together with eigenvalues $\kappa^{(2)}_\alpha$
which lift the degeneracy.

Now we consider the Ritz functional $\mu(\psi)$ (\ref{ritz}) setting
$\psi$ equal to
$\psi_\alpha=\psi^{(0)}_\alpha+g\psi^{(1)}_\alpha+\rO(g^2)$.
Using the relation (\ref{firstorder}) we get
\begin{eqnarray}
 \psi_\alpha^{\dagger} D_WD_W^{\dagger} \psi_\alpha &=&
 (\kappa_\alpha^{(2)})^2g^4 + \rO(g^5) \,,\\
 \psi_\alpha^{\dagger} D_W^{\prime}D_W^{\prime\dagger} \psi_\alpha &=&
 \psi_\alpha^{(0)\dagger} [(D_W^{(1)}-D_W^{\prime(1)})\gamma_5]^2
 \psi_\alpha^{(0)}g^2 + \rO(g^3) \,.
\end{eqnarray}
From (\ref{minritz}) we derive an upper bound for the minimal generalized eigenvalue
$\lambda_{\rm min}$ of (\ref{genEV})
\be
 \lambda_{\rm min} \le {\rm const}\times g^2 \quad (m=0) \,.
\ee
By considering the inverse Ritz functional $\mu^{-1}(\psi)$ and setting
$\psi$ equal to the eigenfunctions $\psi_\alpha^\prime$ of
$D_W^\prime\gamma_5$ we can derive in full analogy to the above
a lower bound for the maximal generalized eigenvalue
$\lambda_{\rm max}$ of (\ref{genEV})
\be
 \lambda_{\rm max} \ge {\rm const}^\prime \times g^{-2} \quad (m=0) \,.
\ee

A detailed analysis of the perturbation expansion of
the generalized eigenvalues themselves, which confirms
the variational arguments given here, is deferred to appendix~\ref{PTapp}.

\subsection{Partially stochastic estimation of determinant ratios}

We saw that a few extremal eigenvalues in $M^\dag M$ can ruin the stochastic
acceptance rate even if the relevant determinant ratio is close to unity.
Moreover, at least in our model, these kind of spectra are really common.
We now develop a mixed strategy treating the bulk of the eigenvalues
stochastically and a few special ones exactly. Of course, detailed balance
has now to be demonstrated for the combination. For the time being we content
ourselves with proofs for $\rho(\eta)=\exp(-\eta^\dag\eta)$ only.
More general cases may well be possible.

We assume, that for each spectrum we can identify a set $S$ such that
$\{\lambda_i | i\in S\}$ consists of the $s=|S|$ 
extremal\footnote{
We assume this construction to be unique, i.~e. no degeneracy at the boundaries 
of $S$.}
eigenvalues of $M^\dag M$ (the same number of large and small ones), 
which we treat deterministically. The associated
eigenvectors $\{ \phi_i| i\in S\}$, that we choose to be orthonormal, 
span an $s$ dimensional subspace that we characterize 
by a projection operator
\be
P = \sum_{i\in S} \phi_i \; \phi_i^\dag \equiv P(A,A')
\ee
Here we remind ourselves that $P$, via the Dirac operators $D_W, D_W'$,
depends on a pair of gauge potentials $A, A'$.
Due to the symmetric inclusion of large and small $\lambda_i$
a spectral analysis of the
inverse operator $(M^\dag M)^{-1}$ with inverted eigenvalues
would lead to the same projector (for the same $M$).

Anticipating a discussion of detailed balance we are interested
in information about the relation between $P(A,A')$ and $P(A',A)$.
As mentioned before, going to the reverse process ($A \leftrightarrow A'$), $M$
changes to $M^{-1}$. Hence in this case we are concerned with the eigenvalue problem of 
$M^{-1\dag}M^{-1}=(MM^\dag)^{-1}$. Its eigen{\it values} are reciprocal to those
of $M^\dag M$ but the eigen{\it vectors} and hence the associated projectors are different,
$P(A',A) \not = P(A,A')$. The problems are however related by a unitary transformation
\be
MM^\dag = U^\dag M^\dag M U, \quad U^\dag U =1= UU^\dag \, .
\ee
Explicitly, U may be written as
\be
U= M^\dag (MM^\dag)^{-1/2} \, .
\ee
The same unitary transformation relates the eigenvectors of the two problems
such that
\be
P(A',A) = U^\dag P(A,A') U \, .
\ee
holds.

Returning to the forward problem we can factorize
\be
M^\dag M = (\bar{P} + P M^\dag M)  (P + \bar{P} M^\dag M)  
\label{Pfactorize}
\ee
where we introduced the complementary projector 
\be
\bar{P} = 1 - P
\ee
and used standard properties of projectors 
and commutativity $PM^\dag M=M^\dag MP$, which is easily seen in the spectral
representation of $M^\dag M$.
While both factors in (\ref{Pfactorize}) are nonsingular operators in the full
domain, they have a block structure with unit operators in the subspaces
of $\bar{P}$ and $P$.
The exact ``small'' determinant of the first factor is
\be
\det(\bar{P} + P M^\dag M) = \prod_{i\in S} \lambda_i .
\ee
A stochastic estimator for an acceptance step with the second factor
alone would be given by (see (\ref{w0stocc}))
\[
\min\left[1,\exp(-\eta^\dag \bar{P}(M^\dag M-1)\bar{P}\eta)\right] \; .
\]
As the true partially stochastic acceptance criterion $w_s(A,A')$
to accept a proposed $A'$ given an `old' $A$ we now propose\footnote{
One could also think of two separate successive Metropolis steps,
but this leads to smaller overall acceptance rates.}
\be
w_s(A,A') = \min\left[1,\prod_{i\in S} \lambda_i^{-1}
\exp(-\eta^\dag \bar{P}(M^\dag M-1)\bar{P}\eta)\right]
\ee
To prove detailed balance we start from
\be
w_s(A',A) = \min\left[1,\prod_{i\in S} \lambda_i
\exp(-\eta^\dag U^\dag \bar{P}U ((M M^\dag)^{-1}-1)U^\dag \bar{P}U \eta)\right] \, .
\ee
In
\bea
&& \langle w_s(A',A) \rangle_\eta = \\[1ex]
&& \int D[\eta]
\min\left[\exp(-\eta^\dag\eta) , \prod_{i\in S} \lambda_i
\exp(-\eta^\dag U^\dag(P+\bar{P}(M^\dag M)^{-1})U\eta)
\right] \nonumber
\eea
a change of variables
\be
\eta = U^\dag (P+\bar{P}M^\dag M)^{1/2} \eta'
\ee
yields the desired result
\be
\frac{\langle w_s(A,A') \rangle_\eta}{\langle w_s(A',A) \rangle_\eta}=
\prod_{i\in S} \lambda_i^{-1} \det(P+\bar{P}M^\dag M)^{-1}=|\det(M)|^{-2} \, .
\ee

With the above formalism we have an algorithm which avoids the low stochastic
acceptance from extremal eigenvalues if their product is of order unity.
It remains to discuss how to compute the
required eigenvalues and -vectors. We do not attempt a detailed discussion here.
An obvious idea is however to generalize the method used by the ALPHA
collaboration in the past
to obtain low and high lying eigenvectors of the ordinary problem. It is
based on minimizing a Ritz functional 
by a conjugate gradient technique \cite{Kalkreuter:1996mm}.
The relevant functional for the generalized problem is
\be
\mu(\chi) = \frac{|(D_W+m)\chi|^2}{|(D_W'+m)\chi|^2}
\ee
This functional is extremal at generalized eigenvectors and its value there
is one of the generalized eigenvalues which coincide with those of $M^\dag M$.
Different generalized eigenvectors $\chi_i$ are not orthogonal, but one may show
that $\chi_i^\dag (D_W+m) (D_W+m)^\dag \chi_j = 0$ if $\lambda_i\not=\lambda_j$.
In fact, $(D_W+m)^\dag \chi_i$ are the (unnormalized) eigenvectors of $M^\dag M$
and $(D_W'+m)^\dag \chi_i$ are those of $MM^\dag$.
After finding the absolute minimum of $\mu$ at $\chi_1$ one may then search
for the minimum in the space orthogonal to $(D_W+m) (D_W+m)^\dag \chi_1$
to find $\chi_2$ belonging to the second smallest eigenvalue $\lambda_2$.
For the largest eigenvalues one may proceed analogously or just exchange
numerator and denominator of $\mu$. Note, that we do not need a solver during
the minimization, that is no intolerable nested iterations. A solver is needed
however in the stochastic part to apply $M$ or $M^\dag M$ to $\bar{P}\eta$. Due to the
filtering of the random vector through the projector this should however be a
well-conditioned problem and take few iterations.
An alternative method to construct the required eigenvalues and vectors
could be Lanczos techniques \cite{Matrix-Computations}
which are known to first converge for the extremal eigenvectors needed here.

\section{Numerical experiments with PSD}

\subsection{Acceptance rate for PSD \label{qforPSD}}

We now consider the
partially stochastic acceptance rate
\be
q_s = \frac1Z \int D\mu(A) |\det(D_W+m)|^2 \int D\mu(A') \;
\langle w_s \rangle_\eta \, .
\label{pstochacc}
\ee
The `observable' $\langle w_s \rangle_\eta$ in this double pathintegral depends
on the generalized eigenvalues $\lambda_i$ and on their division
into the deterministic subset $S$ and the stochastic one $\bar{S}$.
As limiting cases it includes the fully deterministic and fully stochastic 
evaluation.
For a numerical estimation of $q_s$ we produce pairs
of configurations $A_k,A_k'$ 
and determine for all of them, at some value of $z$,
$d_k=|\det(D_W+m)|^2$, generalized eigenvalues $\lambda_i$ and from them
estimates 
$a_k=\langle w_s \rangle_\eta$ 
according to (\ref{Qrhouint}, \ref{Qpartial}).
Then we have the Monte Carlo estimate
\be
q_s \approx \frac{\sum_k d_k a_k}{\sum_k d_k} \, .
\ee
In Fig.\ref{accvss} it is shown how the average acceptance rises if starting
from the fully stochastic case $s=0$ more and more eigenvalues are included in
$S$. Parameters are $L=8, z=1, m=0.0125$ and 1000 pairs of
configurations are sampled. 
\begin{figure}[ht]
 \begin{center}
    \includegraphics[width=12cm]{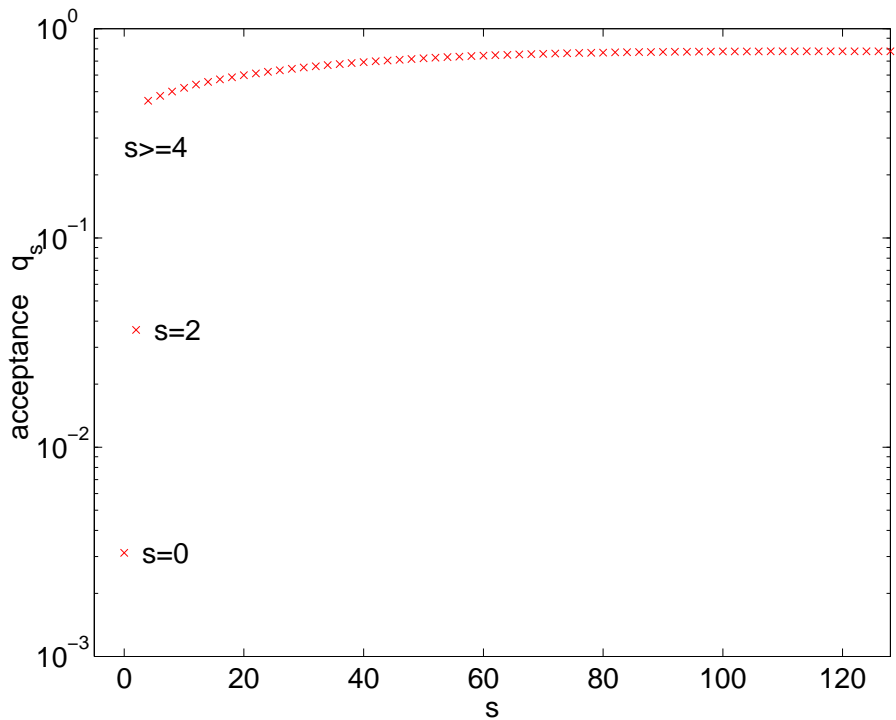}
 \end{center}
 \caption{The acceptance rate vs. the number $s$ of nonstochastic generalized
eigenvalues (parameters B in Tab.\ref{qStab}). 
 \label{accvss}}
\end{figure}
Some further examples can be found in Tab.\ref{qStab}.
\begin{table}[htb] 
 \centering
 \begin{tabular}{|llllllc|}
  \hline
  $L$  & $z$& $~~~m$ & $~~~~q$ & $~~~q_0$ & $~~~q_4$ & label\\[0.5ex]
  \hline\hline
8 & 1 & 0.0250 & 0.837(7)~ &  0.0141(2) & 0.460(4) & A \\
8 & 1 & 0.0125 & 0.734(11) & 0.0029(1) & 0.425(6) & B \\
8 & 1 & 0.0050 & 0.602(15) & 0.00061(2) & 0.368(8) & C \\
8 & 2 & 0.0125 & 0.634(14) & 0.0020(1) & 0.130(3) & D\\
8 & 4 & 0.035  & 0.819(8) & 0.00084(4) & 0.0083(2) & E \\
  \hline 
 \end{tabular} 
\caption{Values for acceptance rates for the deterministic ($q$),
stochastic ($q_0$) and mixed case ($q_4$) with four
extremal eigenvalues in $S$.}
\label{qStab}
\end{table}
We note that there is a roundoff problem in the straightforward
evaluation of (\ref{Qpartial}) due to cancellations and significance loss.
With standard double precision accuracy
some clever recombination of terms would be
required before using the formula much beyond $L=8$ with $n=128$ eigenvalues 
$\lambda_i$.

At this point we also study the dependence of the acceptances 
on the `size' of the proposed moves. Global heatbath steps with respect
to the gauge action that we have used so far are of maximal size.
Lowering from one the parameter $t$ that we have introduced in (\ref{tsteps})   
allows us to mimic smaller steps. 
\begin{figure}[ht]
  \begin{center}
    \includegraphics[width=12cm]{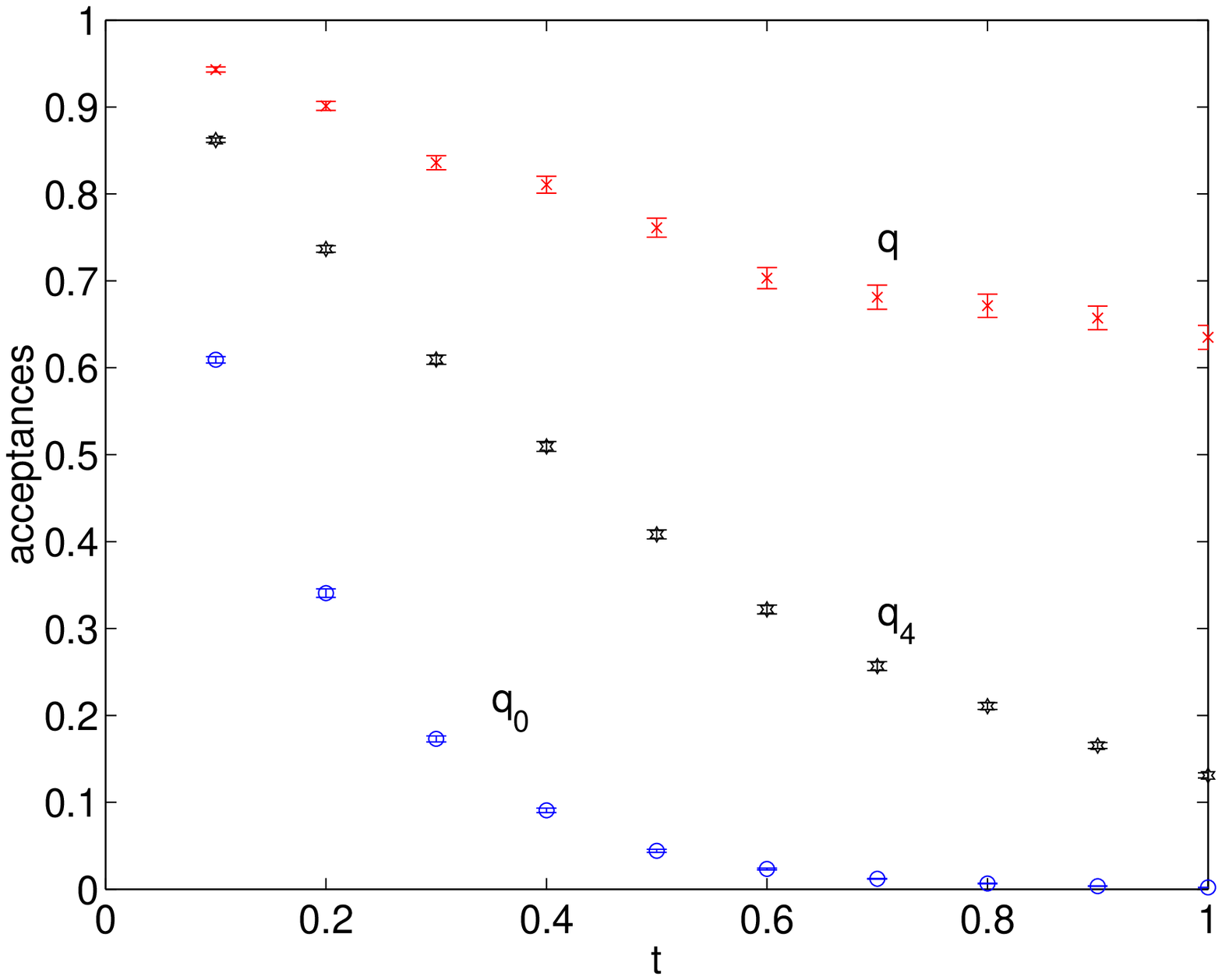}
  \end{center}
\caption{Acceptances rates versus stepsize $t$ for $L=8,m=0.0125,z=2$.}
\label{accvst}
\end{figure}
In Fig.\ref{accvst} we see the expected dependence
of the acceptances. In particular we see that partially stochastic 
acceptance steps allow for much larger moves without excessive rejections.

Also in the partially stochastic case we have found a successful
Gaussian model for the acceptance rate.
Our starting point here is (\ref{pstochacc}) which we break into two steps
\be
\tilde{\nu}_s(\D) = 
\llangle \; \langle \,
\delta[\D-\sum_{i\in S}\ln\lambda_i -\eta^\dag \bar{P} (M^\dag M-1)\eta] 
\, \rangle_\eta \; \rrangle
\label{nutildePSD}
\ee
with
\be
\llangle {\cal O} \rrangle = \frac1Z \int D\mu(A) |\det(D_W+m)|^2 \int D\mu(A')\, {\cal O}
\ee
and
\be
q_s = \int_{-\infty}^{+\infty} d\D \, \tilde{\nu}_s(\D) \, \min(1,\exp(-\D)) .
\ee
As a model we assume $\tilde{\nu}_s(\D)$ to be a Gaussian of mean $m_s$ and width $b_s$.
For it $q_s$ evaluates to
\be\label{qsgauss}
q_s=\frac12 \, \erfc\left(\frac{m_s}{\sqrt{2}b_s}\right) +
    \frac12 \, \erfc\left(\frac{b_s^2-m_s}{\sqrt{2}b_s}\right) \exp(b_s^2/2-m_s) \, .
\ee
In the expressions for the mean and width of the $\D$ distribution the $\eta$
integrations can be performed and we find
\bea
m_s &=& \llangle\, \sum_{i\in S} \ln\lambda_i  + 
                   \sum_{i\in \bar{S}} (\lambda_i-1) \,\rrangle\\
b_s^2 &=& \llangle\, [\, \sum_{i\in S} \ln\lambda_i  + 
                   \sum_{i\in \bar{S}} (\lambda_i-1)\, ]^2 \,\rrangle -m_s^2
      + \llangle\, \sum_{i\in \bar{S}} (\lambda_i-1)^2   \,\rrangle .
\eea
In Tab.\ref{qStab} the last column is replaced by (0.452,0.419,0.364,0.140)
in the Gaussian model for A--D, while the agreement is worse for case E
with its different pattern of fluctuating eigenvalues.

\subsection{The relevance of gaugefixing}

We now turn to the question of gauge fixing. Up to now all gaugefields were
generated in one and the same completely fixed gauge according to 
(\ref{Gaussmeasure}).
Stochastic acceptance
steps depend on the operator $M$ of (\ref{Maccrej}). If only $\rho(\eta)$
is invariant under $x$-dependent phase transformations of the random field 
$\eta(x)$ then all acceptances are invariant under a change of gauge,
that is the {\em same} transformation applied to $A$ and $A'$.
If however the update proposal includes a more or less random gauge move
of $A'$ {\em relative} to $A$, this is equivalent to only gauge transforming
$A'$. In this case the nonstochastic acceptance probability involving
the ratio of gauge invariant
determinants as in (\ref{qdet}) is unchanged. This is however
not true for the eigenvalues $\lambda_i$ of $M^\dag M$ and of the generalized
eigenvalue problem  (\ref{genEV}). We looked at the generalized spectrum
for pairs $A_\mu,A'_\mu+\Delta_\mu \gamma$ for fixed $A,A'$ and a number of
randomly chosen gauge functions $\gamma(x)$. For nonvanishing $\gamma$
the eigenvalues $\lambda_i$ show much more variation and smaller values
of $F(\lambda_i)$ in (\ref{Qrhouint}) result
while the ratio of determinants 
$\prod_i \lambda_i$ remains unchanged as it has to.

A non gauge-fixed simulation is achieved by replacing in (\ref{Gaussmeasure})
\be
\prod_x \delta(\D_\mu^* A_\mu) \to
\exp\left(-\frac1{2\xi^2}\sum_x (\D_\mu^* A_\mu)^2 \right)
\ee
such that the value $\xi=0$ corresponds to the previous case.
It is not difficult to show that a global heatbath step for
this measure is effected by a gaugefixed step as before followed
by a gauge transformation with a gauge function $\gamma=\xi\phi$.
Here a new independent random field $\phi$ is produced
obeying (\ref{phitildenull}, \ref{phitildecov}).
Acceptance rates as functions of the gauge parameter are shown in
Fig.\ref{accvsxi}.
\begin{figure}[ht]
  \begin{center}
    \includegraphics[width=12cm]{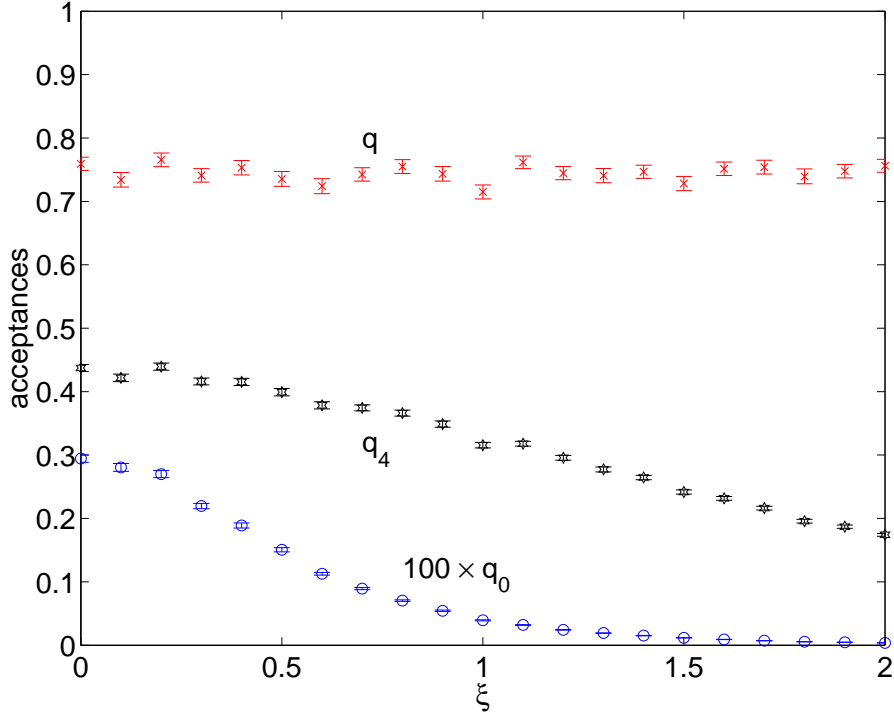}
  \end{center}
\caption{Acceptance rates versus gauge parameter $\xi$ for $L=8,m=0.0125,z=1$.}
\label{accvsxi}
\end{figure}
We notice the dramatic decay of the already small purely stochastic
acceptance.

\subsection{Some unquenched simulations with PSD}

As a practical application
we performed a few simulations on lattices $L=24$ with $z=1$ and $m=0$. By looking
at Tab.\ref{mctab} we observe that for this value of $z$ the value of
$\sigma_c/g^2$ seems to depend only weakly on $L$  whereas $-m_c/g^2$
increases monotonically with $L$. We have good reasons to expect that 
at these parameters no `exceptional' fields 
with zero eigenvalues of the Wilson operator $D_W+m$ 
are proposed.

We compare PSD simulations at $s=4$
with exact determinants (equivalent to $s=2L^2$).
At $s=4$
we employ both global heatbath ($t=1$) proposals
and smaller update moves with the parameter t in (\ref{tsteps}) set to $1/2$.
We measure the `pion'-susceptibility in the unquenched ensemble
\be
 \chi = \frac{1}{L^2}\,
\left\langle \Tr\left[D_W^{-1\,\dag}D_W^{-1}\right] \right\rangle\,,
\ee
where $\Tr$ refers to both Dirac indices and space.
Before integrating out the Grassmann valued fermion fields
$\chi$ can also be interpreted as a susceptibility of
the density
$\bar{\psi}_i(x)\gamma_5\psi_j(x)$ for two flavors $\psi_i, i=1,2$.
We take $\chi$ here as a simple correlation function with significant
contributions at the scale of the meson correlation length.
\begin{table}[htb] 
 \centering
 \begin{tabular}{|cccrc|}
  \hline
  $s$  & $t$ &    $q_s$ & $\chi$\phantom{(12)} & $\tau_{{\rm int},\chi}$ \\[0.5ex]
  \hline\hline
   $2L^2$ & 1   & 0.61 & 983(12)  & 2.4(3)\\
     4    & 1   & 0.24 & 1005(15) & 4.4(8)\\
     4    & 1/2 & 0.50 & 997(14)  & 3.8(5)\\[0.5ex]
  \hline 
 \end{tabular} 
\caption{Results from simulations of 4900 measurements each
on $L=24$ lattices with $z=1$ and $m=0$.}
\label{chitab}
\end{table}

In Tab.\ref{chitab} we list our results for the acceptance rates
and the susceptibility
$\chi$ together with its integrated autocorrelation time $\tau_{\rm int,\chi}$,
which is in units of (global) gaugefield updates.
The stepsizes $t=1/2$ and $t=1$ are equally efficient here, since the smaller
but equally expensive
steps are precisely balanced by the larger acceptance. An optimization
is not attempted in this model study. It will look different, if the stepsize
is also tuned by the number of modified links and not just by the amount of change. 
\begin{figure}[ht]
  \begin{center}
   \includegraphics[width=12cm]{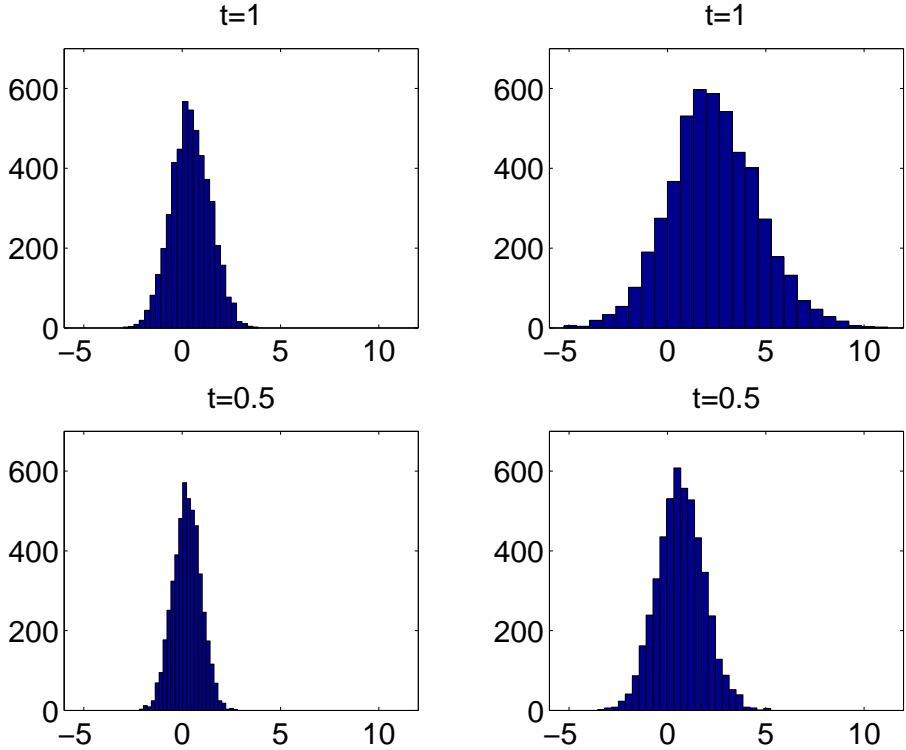}
  \end{center}
\caption{Distribution of $\sum_{i\in S}\ln\lambda_i$ (left) and 
$\eta^\dag \bar{P} (M^\dag M-1)\eta$ (right) contributing to
$\tilde{\nu}$ in (\ref{nutildePSD}).}
\label{delta_hist}
\end{figure}

In the simulations with PSD we also looked at the distribution of the action
difference $\D$ that appears in (\ref{nutildePSD}).
In Fig.\ref{delta_hist} we show histograms with the distributions of the two
components contributing in (\ref{nutildePSD}) 
for the
two choices $t=1$ and $t=0.5$. From
the measured values of $\D$ we extract its mean $m_s$ and variance
$b_s^2$. For $t=1$ we get $m_s=2.82$ and $b_s^2=5.90$, for $t=0.5$ we get
$m_s=0.92$ and $b_s^2=1.88$ both a factor three smaller than for $t=1$.
Inserting these values in (\ref{qsgauss}) we obtain acceptance rates in the
Gaussian model of 24\% for $t=1$ and of 50\% for $t=0.5$, in perfect
agreement with the acceptance rates directly observed in the simulations.

\section{Conclusions}

We found that two-dimensional QED can be simulated
over a wide range of parameters by a combination
of pure gauge update proposals combined with a global
Metropolis step including the exactly evaluated determinant for
two flavours of Wilson fermions. 
Employing a coupling for the proposals that differs from the physical one
can raise the acceptance rate further and is an example of
ultraviolet filtering.
This algorithm provides an upper bound
for the acceptance rate of a class of stochastic techniques.
We studied such methods where we replace
the ideal exact determinant
ratios by less expensive stochastic estimates whose
main cost are inversions of the Dirac operator.

It turned out to be both necessary and possible
to find a mixed form of stochastic estimates combined with the exact
incorporation of a few critical modes which otherwise spoil
the acceptance probability. These modes are given as extremal
eigenvalues of the generalized problem
defined by the Dirac operators in the present and the
newly proposed gaugefield. This was coined PSD algorithm as it is
based on the partially stochastic determinant.
Gauge fixing proved to be advantageous in this case.
If omitted, the update proposals include
random relative moves along the 
gauge orbit which are punished with enhanced rejection.
For the physically smaller volumes
we found the acceptance rates to be well described
by assuming Gaussian distributions for those parts of the total action
that enter the Metropolis decision. The rates are then given
by error functions of the observed mean and variance.
With these insights we plan to investigate similar update schemes
for four dimensional QCD,
especially at small and intermediate physical volume,
and hope to report on this in the future.

{\bf Acknowledgements} We wish to express our thanks to
Burkhard Bunk, Anna Hasenfratz and
Rainer Sommer for helpful discussions, and e-mail exchange 
with Tony Kennedy
is acknowledged. All calculations were carried out
in Matlab employing its built-in linear algebra tools.


\begin{appendix}
\section{Acceptance rate for given spectrum}
\label{Qformula}

In this appendix we evaluate the integral
\be
F(\lambda_i) = \prod_{i\in\bar{S}} \left(\int_0^\infty d u_i \right)
\min\Bigl[\exp(-\sum_{i\in\bar{S}} u_i)\, ,\, 
\prod_{i\in S} \lambda_i^{-1}  \exp(-\sum_{i\in\bar{S}} \lambda_i u_i)\Bigr].
\label{Qrhouint}
\ee
Eigenvalues $0<\lambda_i<\infty, i=1\ldots n$ enter here and $S$ is a subset
of their indices which may also be empty.
The set $\bar{S}$ comprises the remaining indices.

The part, where the first argument of $\min$ is minimal, is given by
\be
f(\lambda_i) = \prod_{i\in\bar{S}} \left(\int_0^\infty d u_i \right) 
\exp(-\sum_{i\in\bar{S}} u_i) \; \theta\Bigl[
\sum_{i\in\bar{S}} (1-\lambda_i) u_i -C \Bigr]
\label{fdef}
\ee
where we abbreviated
\be
C = \sum_{i\in S} \ln\lambda_i \, .
\ee
By rescaling integrals the full result is obtained as
\be
F(\lambda_i) = f(\lambda_i) + (\prod_i\lambda_i^{-1}) f(\lambda_i^{-1}),
\label{Qbyf}
\ee
a form which immediately reflects detailed balance in the form
\be
\frac{F(\lambda_i)}{F(\lambda_i^{-1})}= \prod_i \lambda_i^{-1}
\ee

To evaluate $f$ we use the Fourier representation of the step function
\be
\theta(u) = \int_{-i\infty}^{+i\infty} \frac{dz}{2\pi i} \; \frac{\re^{zu}}{z+\epsilon},
\ee
where $u$ is real, $\epsilon$ positive and infinitesimal, and the integration
runs along the imaginary axis. Proof: residue theorem.
With this inserted into (\ref{fdef}) the $u$-integrations factorize
and can be carried out to yield
\be
f(\lambda_i) = \int_{-i\infty}^{+i\infty} \frac{dz}{2\pi i} \; 
\frac{\re^{-zC}}{z+\epsilon} \; \prod_{j\in\bar{S}} \frac1{1-z(1-\lambda_j)}.
\ee
Assuming non degenerate $\lambda_i$ also this integral is evaluated
by the residue theorem. The contour can be closed with negligible
contribution in the right or in the left half of the complex plane
depending on whether $C$ is positive or negative.
Performing the integration for both cases we find
\be
f(\lambda_i) = \left\{ \begin{array}{rc}
\sum_{i\in\bar{S},\lambda_i<1} \;
\re^{\frac{C}{\lambda_i-1}} \prod_{j\in\bar{S},j\not=i} \; 
\frac{\lambda_i-1}{\lambda_i-\lambda_j} &
\mbox{ for $C>0$}\\[2ex]
1 - \sum_{i\in\bar{S},\lambda_i>1} \;
\re^{\frac{C}{\lambda_i-1}} \prod_{j\in\bar{S},j\not=i} \; 
\frac{\lambda_i-1}{\lambda_i-\lambda_j} &
\mbox{ for $C<0$}
\end{array} \right.
\ee
In the special case $C=0$ either closure of the contours is
legitimate and the two expressions coincide.
The implied identity, which holds for arbitrary
non-coinciding $\lambda_i$,
\be
\sum_{i=1}^n \;(\lambda_i-1)^{n-1}
\prod_{i\not=j=1}^n \; \frac1{\lambda_i-\lambda_j} = 1
\label{sumlam}
\ee
can in fact also be verified purely algebraically.
To this end we recall the Lagrange polynomials,
which are familiar from interpolation and numerical integration,
\be
l_i(\lambda)=\prod_{j\not=i} \frac{\lambda-\lambda_j}{\lambda_i-\lambda_j} \, ,
\ee
with the property
\be
l_i(\lambda_j)=\delta_{ij} \, .
\label{Lpolrel}
\ee
We may also introduce a matrix $A$ with elements $a_{ij}$ by expanding
\be
l_i(\lambda) = \sum_j a_{ij} \lambda^{j-1} \, ,
\ee
and in particular its last column is given by
\be
a_{in} = \prod_{j\not=i} \frac1{\lambda_i-\lambda_j} \, .
\ee
The famous Vandermonde matrix $V$ has matrix elements
\be
v_{jk} = (\lambda_k)^{j-1}
\ee
and in terms of it (\ref{Lpolrel}) says  that $A$ and $V$ are inverse to each other.
A particular consequence is the relation
\be
\sum_i v_{ki} a_{in} = \sum_i (\lambda_i)^{k-1} 
\prod_{j\not=i} \frac1{\lambda_i-\lambda_j} = \delta_{kn}
\ee
which also implies (\ref{sumlam}) upon binomial expansion of the numerator.

It now remains to combine the two contributions to $F$ in (\ref{Qbyf}).
A short calculation leads to the  result
\be
F= \left\{ \begin{array}{rc}
\prod_k\lambda_k^{-1} + \sum_{i\in\bar{S},\lambda_i<1}(1-1/\lambda_i)\,
\re^{\frac{C}{\lambda_i-1}} \prod_{j\in\bar{S},j\not=i} \; 
\frac{\lambda_i-1}{\lambda_i-\lambda_j} &
\mbox{ for $C>0$}\\[2ex]
\label{Qpartial}
1\quad - \sum_{i\in\bar{S},\lambda_i>1}(1-1/\lambda_i)\,
\re^{\frac{C}{\lambda_i-1}} \prod_{j\in\bar{S},j\not=i} \; 
\frac{\lambda_i-1}{\lambda_i-\lambda_j} &
\mbox{ for $C<0$}
\end{array} \right.
\ee
Note that here the exponentials are both damping and may also be written as
\be
\re^{\frac{C}{\lambda_i-1}} = \prod_{k\in S}(\lambda_k)^{\frac1{\lambda_i-1}} \, .
\ee
For $\bar{S}=\emptyset$ we recognize the deterministic acceptance
$\min[1,\prod_k\lambda_k^{-1}]$. For the fully stochastic case $S=\emptyset$
we may use identity (\ref{sumlam}) and its companion
\be
\sum_{i} \; \frac1{\lambda_i} (\lambda_i-1)^{n-1}
\prod_{j\not=i} \; \frac1{\lambda_i-\lambda_j} = \prod_k\lambda_k^{-1}
\ee
to obtain (\ref{Qrhoinlambda}).

\section{Perturbation theory for generalized eigenvalues}
\label{PTapp}
\subsection{General expressions}

We want to solve the generalized eigenvalue problem
\be
H \chi = \lambda H' \chi
\label{geneig}
\ee
by finding $\lambda$ as stationary values of
\be
r(\chi) = \frac{\mu(\chi)}{\mu'(\chi)}, \; \mu=\chi^\dag H \chi, \;
\mu'=\chi^\dag H' \chi .
\ee
The operators $H=h^2,H'=h'^2$ (all hermitian) possess expansions
\bea
h&=&h_0+\sum_{k\ge 1} g^k h_k \\
h'&=&h_0+\sum_{k\ge 1} g^k h'_k
\eea
and similarly for $H, H'$.
We are first interested in the case that $h_0$ has a zero mode
\be
h_0 \phi_0 = 0, \; \phi_0^\dag \phi_0 =1,
\ee
which in perturbation theory is only lifted in second order because we assume
\be
\phi_0^\dag h_1 \phi_0=\phi_0^\dag h'_1 \phi_0=0 .
\ee
In this case there is an eigenvector $\phi = \phi_0 + g \phi_1 +\rO(g^2)$ of $h$,
\be
h \phi = \kappa \phi , \quad \kappa = \kappa_2 g^2 +\rO(g^3) .
\label{kappaex}
\ee
with
\be
h_0\phi_1+h_1\phi_0=0
\ee
and
\be
\kappa_2 = \phi_0^\dag (h_2 - h_1 (h_0)^{-1} h_1)\phi_0 .
\ee
We next want to show that {\em to leading order in $g$ } the vector
$\phi$ is also a stationary point
of $r$ with the corresponding generalized eigenvalue 
$\lambda=r(\phi)(1+\rO(g))$.

We digress here to a simpler case to show the structure of the argument.
Assume we want to know an extremal value of $r(x)=p(x)/q(x)$ close to an
already known extremum $p'(x_*)=0$ with a small value of $p(x_*)/q(x_*)=\epsilon$.
Then we set $x=x_*+y$, expand,  and find that the extremum of $r$ is at
\be
y=\epsilon \, \frac{q'(x_*)}{p''(x_*)} +\rO(\epsilon^2)
\ee
with value $r(x_*+y)=\epsilon+\rO(\epsilon^2)$ unchanged to leading
order of the expansion.

Returning to the real problem we set
\be
\chi = \phi + R, \; R^\dag \phi=0.
\label{Rdef}
\ee
Stationarity of $r$ implies
\be
H \chi \, \mu'(\chi) = H' \chi \, \mu(\chi)
\label{rstationary}
\ee
It is not too difficult to see that with our Ansatz we get
\bea
\mu(\chi) &=& g^4 \kappa_2^2 +\rO(g^6) + |hR|^2 \\
h' \chi &=& g (h'_1-h_1) \phi_0 +\rO(g^2) + h' R
\eea
There is a solution to the above equations of the form
\be
R = g^3 \sum_{k\ge 0}g^k R_k.
\ee
This implies $\mu(\chi)=\mu(\phi) + \rO(g^6)$, $\mu'(\chi)=\mu'(\phi) + \rO(g^4)$
and
\bea
\lambda &=& \lambda_2 g^2 + \rO(g^3) = r(\phi) + \rO(g^3)\\
\lambda_2 &=&\frac{(\kappa_2)^2}{|(h'_1 - h_1) \phi_0|^2}
\label{lambda2}
\eea
Using (\ref{rstationary})
the leading term $R_0$ is determined by
\be
h_0 R_0 = (h'_1 - h_1) \phi_0 \, \lambda_2 , \phi_0^\dag R_0=0
\ee
with the last condition following from (\ref{Rdef}).

There is another generalized eigenvalue associated with the
smallest eigenvalue of $h'$. Since the problem has the exact symmetry
$H\leftrightarrow H', \lambda \leftrightarrow 1/\lambda$ it is clear
that the corresponding eigenvalue is 
$\lambda'={\lambda'}_{-2} g^{-2} + \rO(g^{-1})$ 
with
\be
\lambda'_{-2} = \frac{|(h'_1 - h_1) \phi_0|^2}{(\kappa'_2)^2}
\ee
where $\kappa'_2$ is as $\kappa_2$
in (\ref{kappaex}) but now referring to the lowest eigenvalue of
$h'$. The
respective
shift $R'$ of the eigenvector $\phi'$ has a leading component 
$(\kappa_2)^2 R'_0=-(\kappa'_2)^2 R_0$.

We now come to the regular generalized eigenvalues associated with the
non-nullspace of $h_0$, i.e. with zeroth order different from $\phi_0$.
We expand
\bea
\chi &=& \sum_{k\ge 0} g^k \chi_k, \quad h_0 \chi_0 \not= 0 \\
\lambda &=& 1+\sum_{k\ge 1} g^k \lambda_k
\eea
and insert this into (\ref{geneig}) to derive for the first two
nontrivial orders
\bea
(H_1-H'_1) \chi_0 &=& \lambda_1 H_0  \, \chi_0\label{firstorderH}\\
(H_1-H'_1-\lambda_1H_0) \chi_1 &=& 
-(H_2-H'_2-\lambda_2 H_0-\lambda_1 H'_1)  \chi_0 \, .
\eea
Note that the zeroth order is fully degenerate and $\chi_0$ is determined
only in first order by again a generalized eigenvalue problem. This is
similar to ordinary highly degenerate perturbation theory, where a 
large diagonalization cannot be avoided. Combining the first two orders
gives also
\be
\lambda_2 = \frac{\chi_0^\dag(H_2-H'_2-\lambda_1 H'_1)\chi_0}{\chi_0^\dag H_0\chi_0}.
\ee

Two generalized eigenvectors $\chi, \tilde{\chi}$ of (\ref{geneig}) belonging to
different generalized eigenvalues $\lambda\not= \tilde{\lambda}$ are orthogonal in
the sense
\be
0=\chi^\dag H \tilde{\chi} = \chi^\dag H' \tilde{\chi} .
\ee
In particular, eigenvectors to be constructed now have
to be orthogonal to $H$ or $H'$ times the two previous ones.
Of course, this has to come out automatically, but it is interesting to verify
as a consistency check. Using the coincidence with eigenvectors $\phi, \phi'$
up to terms of O($g^3$), the space, to which $\chi_0$ must be orthogonal, may be spanned
by the leading orders of $H\phi'$ and $H'\phi$. A little calculation in ordinary
perturbation theory gives
\bea
H\phi' - H'\phi &=& 2 g (H_1-H'_1) \phi_0 +\rO(g^2) \\
H\phi' + H'\phi &=& g^2 (H_1-H'_1)H_0^{-1}(H_1-H'_1) \phi_0 +\rO(g^3) .
\eea
The orthogonality of solutions $\chi_0$ to these directions follows indeed by
first projecting both sides of
(\ref{firstorderH}) on $\phi_0^\dag$ and then on
$\phi_0^\dag (H_1-H'_1)H_0^{-1}$ and using (\ref{firstorderH}) again.

In our real problem we have to deal with the additional spin degree of
freedom. Hence there are two zero-modes $\phi_{0\alpha}, \alpha=1,2$ of
$h_0$. These are chosen orthonormal and
such that the $2\times 2$ matrix
\be
\phi_{0\alpha}^\dag (h_2 - h_1 (h_0)^{-1} h_1)\phi_{0\beta} = 
\kappa_{2\alpha} \; \delta_{\alpha\beta}
\label{eig22}
\ee
is diagonal. The expansions built upon them generate two
generalized eigenvalues which are O($g^2$) and two O($g^{-2}$).

\subsection{Evaluation for given gaugefields}

Here we evaluate $\kappa_2, \lambda_2, \lambda'_{-2}$ for
gaugefields
\be
A_\mu(x) = \frac1L \sum_p \tilde{A}_\mu(p) \exp(ip\cdot x)
\ee
and an analogous expression for $A'$.
According to (\ref{Afromphi}) we have (for gauge parameter $\xi=0$)
\be
\tilde{A}_\mu(p) = \sum_\nu \epsilon_{\mu\nu} (1-\re^{-ip_\nu}) \tilde{\phi}_r(p) 
\ee
with
\be
\tilde{\phi}_r(p) = \frac12 (\tilde{\phi}(p)+\tilde{\phi}^*(-p))
\ee
leading to a real $A_\mu$.

We consider the expansion of $h=D_W\gamma_5$ now in the ON-basis of free field
states $(1,0) \re^{ipx}/L$ 
and $(0,1)\re^{ipx}/L$, i.e. as kernels in momentum
space and $2\times 2$ matrices in spin. With 
$\hat{p}_\mu = 2\sin(p_\mu/2), \stackrel{\circ}{p}_\mu=\sin(p_\mu)$ we find
\bea
h_0(p,q) &=& \left[
i \sum_\mu \gamma_\mu  \stackrel{\circ}{p}_\mu + \hat{p}^2/2
\right]\gamma_5 \delta_{pq}\\
h_1(p,q) &=& \frac{i}{2L}\sum_\mu \left[ (\gamma_\mu +1)\re^{-ip_\mu} + 
(\gamma_\mu -1)\re^{iq_\mu}
\right]\gamma_5 \tilde{A}_\mu(p-q)\\
h_2(p,q) &=&\frac{1}{4L^2}\sum_\mu \left[ (\gamma_\mu +1)\re^{-ip_\mu} - 
(\gamma_\mu -1)\re^{iq_\mu}
\right]\gamma_5 \widetilde{A^2}_\mu(p-q)
\eea
with $\widetilde{A^2}_\mu(p-q)=\sum_r \tilde{A}_\mu(p-r) \tilde{A}_\mu(r-q)$.
Combinations relevant for (\ref{eig22}) are
\bea
h_2(0,0)&=&
= \frac{\gamma_5}{2L^2} \sum_p \hat{p}^2 \, |\tilde{\phi}_r(p)|^2\\
h_1(p,0)&=&
\frac{1}{2L} \sum_{\mu} \gamma_\mu z_\mu(p) \, \tilde{\phi}_r(p)
\eea
with
\be
(z_0 , z_1) = -2 i \re^{-i(p_0+p_1)/2} \; \Bigl(\hat{p}_0\cos(p_1/2) \, , \,
\hat{p}_1\cos(p_0/2)\Bigr) 
\ee
With some calculation we are then able to show
\bea
&& h_1(0,p)(h_0(p,p))^{-1}h_1(p,0) = \\
&&\frac{-1}{4L^2} \; \left[
\hat{p}^2(2\hat{p}^2 -\hat{p}_0^2\hat{p}_1^2)
+i \sum_\mu \gamma_\mu\stackrel{\circ}{p}_\mu
(4\hat{p}^2-2\sum_\nu|\epsilon_{\mu\nu}|\, \hat{p}_{\nu}^4)
\right]\;
\gamma_5 \frac{|\tilde{\phi}_r(p)|^2}{\stackrel{\circ}{p}^2+\frac14 (\hat{p}^2)^2} \, .
\nonumber
\eea
Upon $p$-summation the part odd in $p$ does not contribute and
from (\ref{eig22}) we get
\be
\kappa_{22} = - \kappa_{21} = \frac1{2L^2} \sum_p
\left[1+
\frac{\hat{p}^2 -\frac12 \hat{p}_0^2\hat{p}_1^2}
{\stackrel{\circ}{p}^2+\frac14 (\hat{p}^2)^2} \, \right]
\hat{p}^2 \, |\tilde{\phi}_r(p)|^2
\ee
Inspecting (\ref{phitildecov}) we find on average in the quenched ensemble
\be
\langle\; |\tilde{\phi}_r(p)|^2 \; \rangle_G = \frac1{(\hat{p}^2)^2}.
\ee
A numerical asymptotic expansion gives
\be
\langle\; \kappa_{22} \; \rangle_G = \frac1{2\pi} \ln L
-0.00639327 + 0.13303 \, L^{-2}
+ \rO(L^{-4}) .
\ee
Also the mean of the square can be worked out
\be
\langle\; (\kappa_{2\alpha})^2 \; \rangle_G =
\langle\; \kappa_{2\alpha} \; \rangle_G^2
+ 0.00773389 + 0.03112 \, L^{-2}
+ \rO(L^{-4}) .
\ee

The denominator in (\ref{lambda2}) is given by elements of the $2\times 2$ matrix
\bea
&&\sum_p (h'_1(p,0)-h_1(p,0))^\dag (h'_1(p,0)-h_1(p,0)) =\\
&&\frac1{L^2}\sum_p (\hat{p}^2 -\frac12 \hat{p}_0^2\hat{p}_1^2) \, 
|\tilde{\phi}'_r(p)-\tilde{\phi}_r(p)|^2 \nonumber
\eea
and produces
\be
\langle\; |(h'_1 - h_1) \phi_{0\alpha}|^2 \; \rangle_G = \frac1{\pi} \ln L 
-0.06162368 + 0.26607 \, L^{-2} + \rO(L^{-4}).
\ee

\end{appendix}

\bibliography{lattice}           

\begin{thebibliography}{10}

\bibitem{Creutz-Book}
M. Creutz,
\newblock Quantum Fields On The Computer (Advanced Series on Directions in High
  Energy Physics --- Vol. 11, World Scientific, Singapore, 1992).

\bibitem{Duane:1987de}
S. Duane, A.D. Kennedy, B.J. Pendleton and D. Roweth,
\newblock Phys. Lett. B195 (1987) 216.

\bibitem{Gottlieb:1987mq}
S. Gottlieb, W. Liu, D. Toussaint, R.L. Renken and R.L. Sugar,
\newblock Phys. Rev. D35 (1987) 2531.

\bibitem{Orginos:1999cr}
MILC, K. Orginos, D. Toussaint and R.L. Sugar,
\newblock Phys. Rev. D60 (1999) 054503, hep-lat/9903032.

\bibitem{Hasenfratz:2001hp}
A. Hasenfratz and F. Knechtli,
\newblock Phys. Rev. D64 (2001) 034504, hep-lat/0103029.

\bibitem{Hasenfratz:2000qb}
P. Hasenfratz et~al.,
\newblock Nucl. Phys. Proc. Suppl. 94 (2001) 627, hep-lat/0010061.

\bibitem{Bietenholz:2000iy}
W. Bietenholz,
\newblock (2000), hep-lat/0007017.

\bibitem{DeGrand:2000tf}
MILC, T. DeGrand,
\newblock Phys. Rev. D63 (2001) 034503, hep-lat/0007046.

\bibitem{Neuberger:1998fp}
H. Neuberger,
\newblock Phys. Lett. B417 (1998) 141, hep-lat/9707022.

\bibitem{Bode:2001jv}
ALPHA, A. Bode et~al.,
\newblock Phys. Lett. B515 (2001) 49, hep-lat/0105003.

\bibitem{Knechtli:2002vp}
ALPHA, F. Knechtli et~al.,
\newblock (2002), hep-lat/0209025.

\bibitem{Lang:1998tm}
C.B. Lang,
\newblock (1998), hep-lat/9907017.

\bibitem{Gattringer:1997qc}
C.R. Gattringer, I. Hip and C.B. Lang,
\newblock Nucl. Phys. B508 (1997) 329, hep-lat/9707011.

\bibitem{Bardeen:1997bp}
W. Bardeen, A. Duncan, E. Eichten and H. Thacker,
\newblock (1997), hep-lat/9705002.

\bibitem{Narayanan:1995sv}
R. Narayanan, H. Neuberger and P. Vranas,
\newblock Phys. Lett. B353 (1995) 507, hep-lat/9503013.

\bibitem{Joo:2001bz}
B. Joo, I. Horvath and K.F. Liu,
\newblock (2001), hep-lat/0112033.

\bibitem{Hasenfratz:2002jn}
A. Hasenfratz and F. Knechtli,
\newblock Comput. Phys. Commun. 148 (2002) 81, hep-lat/0203010.

\bibitem{Hasenbusch:1998yb}
M. Hasenbusch,
\newblock Phys. Rev. D59 (1999) 054505, hep-lat/9807031.

\bibitem{Numerical-Recipes}
W.H. Press, B.P. Flannery, S.A. Teukolsky and W.T. Vetterling,
\newblock Numerical Recipes: The Art of Scientific Computing, 2nd ed.
  (Cambridge University Press, Cambridge (UK) and New York, 1992).

\bibitem{deForcrand:1998sv}
P. de~Forcrand,
\newblock Nucl. Phys. Proc. Suppl. 73 (1999) 822, hep-lat/9809145.

\bibitem{Duncan:1999xh}
A. Duncan, E. Eichten, R. Roskies and H. Thacker,
\newblock Phys. Rev. D60 (1999) 054505, hep-lat/9902015.

\bibitem{Hasenfratz:2002pt}
A. Hasenfratz and A. Alexandru,
\newblock (2002), hep-lat/0209071.

\bibitem{Hasenfratz:2002vv}
A. Hasenfratz,
\newblock (2002), hep-lat/0211007.

\bibitem{Frezzotti:1998yp}
R. Frezzotti and K. Jansen,
\newblock Nucl. Phys. B555 (1999) 432, hep-lat/9808038.

\bibitem{Frezzotti:1998eu}
R. Frezzotti and K. Jansen,
\newblock Nucl. Phys. B555 (1999) 395, hep-lat/9808011.

\bibitem{Kalkreuter:1996mm}
T. Kalkreuter and H. Simma,
\newblock Comput. Phys. Commun. 93 (1996) 33, hep-lat/9507023.

\bibitem{Matrix-Computations}
G.H. Golub and C.F.V. Loan,
\newblock Matrix Computations, 3rd ed. (The Johns Hopkins University Press,
  Baltimore and London, 1996).

\end{thebibliography}
\bibliographystyle{h-elsevier}   

\end{document}